\documentclass[twocolumn,trackchanges]{aastex63}

\usepackage{color}
\usepackage{hyperref}
\hypersetup{
 colorlinks,
 citecolor=blue,
 linkcolor=red,
 urlcolor=blue}

\received{September 29, 2020}
\revised{October 9, 2020}
\accepted{October 27, 2020}

\submitjournal{ApJ}

\shorttitle{Formation of an atypical sunspot light bridge from flux emergence}
\shortauthors{Louis, Beck, \& Choudhary}

\begin{document}

\title{The formation of an atypical sunspot light bridge as a result of large-scale flux emergence}

\correspondingauthor{Rohan Louis}
\email{rlouis@prl.res.in}

\author[0000-0001-5963-8293]{Rohan E. Louis}
\affiliation{Udaipur Solar Observatory, Physical Research Laboratory\\
Dewali Badi Road, Udaipur - 313001, Rajasthan, India}

\author[0000-0001-7706-4158]{Christian Beck}
\affiliation{National Solar Observatory (NSO), \\
3665 Discovery Drive, Boulder, CO 80303, USA}

\author[0000-0002-9308-3639]{Debi P. Choudhary}
\affiliation{Department of Physics and Astronomy, \\
California State University, Northridge (CSUN), CA 91330-8268, USA}

\begin{abstract}
We use a combination of full-disk data from the Solar Dynamics Observatory and high-resolution data from the 
Dunn Solar Telescope (DST) to study the formation, structure, and evolution of an atypical light bridge (LB) 
in a regular sunspot. The LB results from the emergence of magnetic flux with one footpoint rooted in a pore 
outside the parent sunspot that appears about 17\,hrs before the LB. The pore has a polarity opposite to
that of the sunspot and recedes away from it at a speed of about 0.4\,km\,s$^{-1}$.
This is accompanied by the development of an elongated magnetic channel in the outer penumbra which triggers 
the formation of the LB when it reaches the inner penumbral boundary. The LB is a nearly horizontal structure 
with a field strength of about 1.2\,kG that exhibits long-lived photospheric blue-shifts of about 
0.85\,km\,s$^{-1}$ along its entire length.
The emergence of the LB leads to dynamic surges in the chromosphere and transition region 
about 13\,min later. We derived the photospheric and chromospheric structure of the LB in the DST data from 
spectral line parameters and inversions of \ion{He}{1} at 1083\,nm, \ion{Si}{1} at 1082.7\,nm, \ion{Ca}{2} IR at 
854\,nm and H$_\alpha$ at 656\,nm, and speckle-reconstructed imaging at 700\,nm and 430\,nm. The LB shows 
an elongated filamentary shape in the photosphere without lateral extrusions. The thermal inversion of 
\ion{Ca}{2} IR reveals the LB to be about 600--800\,K hotter than the umbra. Different sections of 
the LB are elevated to heights between 400 and 700\,km. Our results indicate that the LB formation is part 
of a flux emergence event with the LB envelope reaching a height of about 29\,Mm before dissolving after 
about 13\,hr. We suggest that the existence of persistent, large-scale photospheric blue-shifts in LBs is the 
most likely criterion to distinguish between flux emergence events and overturning convection in field-free 
umbral intrusions.
\end{abstract}

\keywords{Sunspots(1653) --- Solar magnetic flux emergence(2000) --- Solar magnetic fields (1503) --- 
Solar photosphere(1518) --- Solar chromosphere (1479) --- Solar corona (1483)}

\section{Introduction}
\label{intro}
Sunspots are the strongest and largest concentrations of the magnetic field in the 
solar photosphere \citep{solanki2003}. During their lifetime, the umbral core often
comprises one or more elongated, bright structures called light bridges (LBs). 
LBs are usually present during the early stages of sunspot formation 
\citep{2010A&A...512L...1S} or late stages of sunspot decay 
\citep{1987SoPh..112...49G} and can have an umbral, penumbral, or 
granular morphology depending on their evolutionary epoch 
\citep{1979SoPh...61..297M,2007PASJ...59S.577K,2012ApJ...755...16L}.
High-resolution observations of granular LBs  typically show a dark lane 
running along their central axis \citep{1994ApJ...426..404S,2003ApJ...589L.117B,
2004SoPh..221...65L,2014A&A...568A..60L}, while penumbral LBs exhibit small-scale 
barbs near the edges of the filamentary structure
\citep{2008ApJ...672..684R,2008SoPh..252...43L}.

The magnetic field in LBs is usually weaker and more inclined
in comparison to the neighbouring umbra \citep{1995A&A...302..543R,
1991ApJ...373..683L,1997ApJ...484..900L}. At photospheric heights, LBs can be
perceived as field-free intrusions of hot plasma into the gappy
umbral magnetic field \citep{1979ApJ...234..333P,1986ApJ...302..809C} or 
large-scale magneto-convective structures \citep{1997ApJ...490..458R,2004ApJ...604..906R}.
This intrusion of hot weakly magnetised plasma forces the adjacent umbral magnetic field
to form a canopy above the LB \citep{2006A&A...453.1079J}.
Such a magnetic topology has been suggested to be responsible for a wide variety of 
transient phenomena such as surges in H$_\alpha$ \citep{1973SoPh...28...95R,
2001ApJ...555L..65A,2016A&A...590A..57R}, strong brightenings
and ejections observed in \ion{Ca}{2} H 
\citep{2008SoPh..252...43L,2009ApJ...704L..29L,2009ApJ...696L..66S}, 
small-scale jets \citep{2014A&A...567A..96L,2018ApJ...854...92T}
or brightness enhancements in the transition region \citep{2003ApJ...589L.117B}.

\cite{2007PASJ...59S.577K} reported that the formation of a LB was 
accompanied by several umbral dots 
\citep[UDs;][]{1979A&A....79..128L,1997ApJ...490..458R,1997A&A...328..682S,1997A&A...328..689S} 
emerging from the leading edges of
penumbral filaments, which were seen to rapidly ingress into the umbra. Using 
high resolution observations from the Solar Optical Telescope 
\citep{2008SoPh..249..167T} on board Hinode \citep{2007SoPh..243....3K} spanning several days, 
they identified the precursor of the LB formation to be relatively slow inward moving UDs, 
which emerged well inside the umbra. They concluded that the appearance of the UDs in the central 
part of the umbra led to a weakening of the umbral magnetic field by hot sub-photospheric plasma that
facilitated the emergence of a buoyant flux tube in the form of the intruding penumbral
filament which ultimately formed the LB.
 
The presence of a LB during the early stages of sunspot or active region (AR) formation can be attributed 
to weakly magnetised plasma being squeezed by the horizontal convergence of stronger 
emergent flux \citep{2008ApJ...687.1373C,2010ApJ...720..233C}. In such a case, the LB exhibits 
a large-scale convective upflow that continuously transports horizontal, but mostly weaker,
magnetic field to the surface. This horizontal magnetic field reconnects with the adjacent 
vertical umbral fields, resulting in episodic and intermittent brightenings and surge 
ejections \citep{2015ApJ...811..137T,2015ApJ...811..138T}. While LBs represent large-scale 
convective upflows, they can also be sites of small-scale magnetic flux emergence in the form of 
flat horizontal loops harbouring a siphon flow \citep{2015A&A...584A...1L}.
The interaction of convective flows in the LB with the strongly magnetized environment of the 
sunspot can produce several small-scale magnetic and plasma inhomogeneities in the former
\citep{2009ApJ...704L..29L,2015AdSpR..56.2305L}.

In this article, we study the formation and properties of an unusual LB using satellite-based 
full-disk data from the Solar Dynamics Observatory (SDO) and ground-based high-resolution observations 
from the Dunn Solar Telescope (DST). Sections \ref{obs} and \ref{secana} describe the data used and the 
analysis techniques, respectively. We present our results in Section \ref{res}. The discussion and 
conclusions are included in Sections \ref{discuss} and \ref{conclu}, respectively.

\section{Observations}
\label{obs}
We study the leading sunspot in NOAA AR 12002 from 2014 March 12 23:00 UT to 06:20 UT 
of March 14. The long-term evolution of the AR is analysed using SDO observations while 
a complex instrument suite at the DST is employed to study 
the magnetic and thermodynamic properties of the sunspot with high resolution at a particular
epoch of its evolution on March 13 from about 20:37--21:00 UT. Both data sets are described below.

\begin{table*}[!ht]
\caption{Characteristics of the DST observations.} \label{tab1}
\centering
\begin{tabular}{c|ccc|cc|c|c}
instrument & \multicolumn{3}{c|}{SPINOR} & \multicolumn{2}{c|}{IBIS} & UBF & G-band \cr\hline
$\lambda$ [nm] & 1082$\pm2$  & 854$\pm2$ & 656$\pm1$  &  854.2$\pm0.2$  & 656.25$\pm0.2$ & 656 & 430 \cr
spatial x/y [$^{\prime\prime}$\,pixel$^{-1}$] & 0.36/0.561 &  0.36/0.366 &  0.36/0.359 & 
\multicolumn{2}{c|}{(0.098)$^2$} & (0.139)$^2$  & (0.085)$^2$  \cr
spectral [pm\,pixel$^{-1}$] & 9.29 & 5.85 & 4.19 &  \multicolumn{2}{c|}{$\sim$ 4 -- 40} & -- & -- \cr
FOV  x-y [$^{\prime\prime}$] & \multicolumn{3}{c|}{60 -- 110} &  \multicolumn{2}{c|}{(90)$^2$} & (88)$^2$&  
(140)$^2$ \cr
exposure time [ms]& \multicolumn{3}{c|}{100} &\multicolumn{2}{c|}{80} & 200 & 10\cr
cadence [sec] & \multicolumn{3}{c|}{--} & \multicolumn{2}{c|}{12} & 7 & 7 \cr
\end{tabular}
\end{table*}

\subsection{SDO Data}
\label{sdodata}
The SDO data consists of continuum intensity filtergrams, Dopplergrams and 
the vector magnetic field from NASA's Helioseismic Magnetic Imager \citep[HMI;][]
{2012SoPh..275..229S} at a cadence of 12\,min. The vector magnetic field products have 
been projected and remapped to a cylindrical equal area Cartesian coordinate system 
centred on the AR. In addition, we utilize similar data at a cadence of 
1\,hr between 13:00 UT and 23:00 UT of March 12 and from 09:12 UT to 23:00 UT
of March 14. The HMI data were complemented with images from the Atmospheric Imaging 
Assembly \citep[AIA;][]{2012SoPh..275...17L} at a cadence of 5\,min in the 1700\,\AA, 
304\,\AA, and 171\,\AA~channels. HMI data gaps occurred between 06:12 UT to 09:12 UT on 
March 13 and March 14. A similar data gap was also present in the AIA 
time series from 06:20 UT to 07:35 UT on March 13 and March 14.

\subsection{High-resolution DST Data} 
\label{dst}
\begin{figure}
\resizebox{8.2cm}{!}{\includegraphics{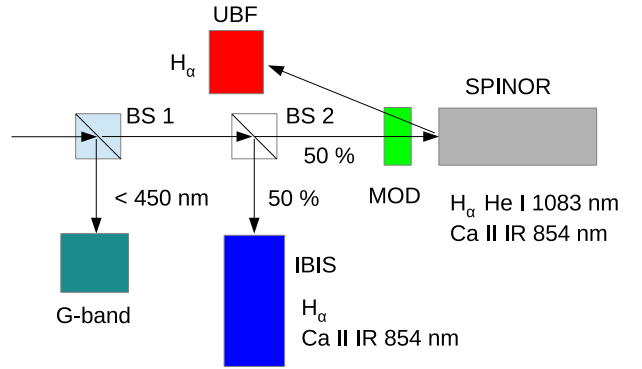}}
\caption{Schematical drawing of the setup at the DST. Beam splitter 1 (BS 1) reflected all light below 
450\,nm towards a G-band imaging camera at 430\,nm. The 50-50 achromatic BS 2 split the light 
evenly between IBIS (H$_\alpha$, \ion{Ca}{2} IR 854\,nm) and SPINOR (H$_\alpha$, \ion{Ca}{2} 
IR 854\,nm, \ion{He}{1} 1083\,nm). The light reflected by the slit jaw passed through the UBF 
that was tuned to the line core of H$_\alpha$. This beam passed twice through the rotating 
modulator (MOD).}
\label{fig_setup}
\end{figure}

For the DST observations, we used a combination of the Interferometric BI-dimensional Spectrometer 
\citep[IBIS;][]{cavallini2006,reardon+cavallini2008}, the SPectropolarimeter for Infrared and 
Optical Regions \citep[SPINOR;][]{socasnavarro+etal2006}, the Universal Birefringent Filter 
\citep[UBF;][]{beckers+etal1975} and one more imaging camera. A 
sketch of the setup is shown in Figure \ref{fig_setup}. A dichroic beam splitter (BS) was 
used to reflect all light below 450\,nm towards a G-band imager. A second, achromatic 50-50 BS 
split the rest of the light evenly between IBIS and SPINOR. The modulator of SPINOR, 
a rotating wave plate, was placed directly in front of its entrance slit. The wave plate 
unfortunately has a small wedge angle that causes a beam wobble with a cone angle of about 
0.003\,deg. With the placement as close to the slit plane as possible, the beam wobble has 
a diameter of about 0\farcs14 on the Sun. The physical slit width of SPINOR of 0\farcs22 on 
the Sun was therefore increased to an effective slit width of 0\farcs36. 
The SPINOR slit-jaw (SJ) unit was reflecting light towards a second 
imaging channel through the UBF that was tuned to the line core of H$_\alpha$. There is no 
image motion of the solar image because of the double passage through the rotating modulator, 
but the slit plane, and hence the slit, the hairlines and all dust particles on its surface 
exhibit a circular image motion in the H$_\alpha$ imaging channel. In addition, the spatial 
scanning of the solar surface for SPINOR is done by moving the slit unit. It was thus not
possible to obtain a complete gain correction for the  H$_\alpha$ SJ images. 

With IBIS, we sampled the chromospheric spectral lines of \ion{Ca}{2} IR at 854.2\,nm and H$_\alpha$
at 656\,nm in spectroscopic mode. With SPINOR, we obtained Stokes vector polarimetry of \ion{He}{1} at 
1083\,nm and \ion{Ca}{2} IR at 854\,nm. A third camera for observing H$_\alpha$ with SPINOR turned out 
to be not correctly synchronized to the modulator position on that day, so that only spectroscopic 
data was acquired. The spatial, spectral and temporal sampling of all instruments is listed in 
Table \ref{tab1}. The seeing conditions were only medium; the data originally were only intended 
as performance test of the rather complex setup.

With the setup described above, we observed the leading spot of NOAA 12002 on 
March 13, 2014, from about UT 20:37 to UT 21:00. The sunspot was located at $x, y \sim 40^{\prime\prime}, 
-180^{\prime\prime}$ at a heliocentric angle of about 12$^\circ$. We scanned the field-of-view (FOV) with SPINOR with 
200 steps of 0\farcs293 step width with an integration time of 5\,s per step. The G-band and H$_\alpha$ 
imagers were triggered by SPINOR at the start of each scan step. 
IBIS sequentially scanned \ion{Ca}{2} IR 854\,nm and H$_\alpha$ with a nonequidistant sampling of 27 
wavelength points for each line and obtained 63 spectral scans of the two spectral lines between UT 
20:44 and UT 20:57 at a cadence of 12 seconds. Table \ref{tab1} lists other technical characteristics 
of the DST data.

\section{Data Analysis}\label{secana}
\subsection{Speckle Reconstruction}
We used the Kiepenheuer-Institut Speckle Interferometry Package \citep{woeger+vdl2008,woeger+etal2008} 
to run a speckle reconstruction of the IBIS broad-band (BB) images at 700\,nm, the G-band images at 
430\,nm and the H$_\alpha$ SJ images at 656\,nm. For the former, we combined the 54 BB images of each 
spectral scan acquired simultaneously with the spectra into one speckle burst for a total of 63 bursts. 
For the latter two, we only used the last 60 of the 200 images in two speckle bursts of 30 images each. 
Some of the IBIS bursts got impacted by clouds, while for the H$_\alpha$ images the reconstruction amplified 
the flat-field residuals caused by the beam wobble of the modulator. We thus dropped the UBF H$_\alpha$ 
images from further analysis and maintained only the second G-band reconstruction and the three IBIS BB 
reconstructions of the spectral scans Nos.~2, 38 and 61 at times of good seeing.

\subsection{Line Parameters}
\paragraph{Spectroscopy} 
For all intensity spectra, we determined the following quantities: the continuum or line-wing intensity, 
the line-core intensity for chromospheric lines, the line-core velocity of all chromospheric lines and 
of the photospheric \ion{Si}{1} line at 1082.7\,nm, and bisectors at either 10 or 30 equidistant line depth 
levels between the line core and the continuum or line wing for all chromospheric lines. We stored the 
intensity value of the bisector level, its length, i.e., the line width at that level, 
and the central position \citep[see, e.g.,][their Figure 3]{2020ApJ...891..119B}.

\paragraph{Polarimetry}
For all polarimetric data, we determined the maximum polarization degree 
$p = \mbox{max}\,\left(\sqrt{Q^2+U^2+V^2}/I (\lambda) \right)$ and the maximum linear polarization 
$L = \mbox{max}\,\left(\sqrt{Q^2+U^2}/I (\lambda)\right)$. The polarization degree $p$ indicates the 
presence of solar magnetic fields through the amplitude of the polarization signals created by the Zeeman effect. Its value 
was used as a threshold in the explicit inversion of spectra to distinguish field-free and magnetic locations 
\citep[e.g.,][]{2007A&A...472..607B}. The linear polarization $L$ is proportional to the modulus of the transverse 
magnetic flux density \citep{2008ApJ...672.1237L} and indicates a magnetic field vector that is inclined to the 
line-of-sight (LOS).

\subsection{Inversions of Spectra}
The Zeeman-sensitive \ion{Si}{1} line at 1082.7\,nm in the SPINOR data was inverted with the Stokes Inversion 
based on Response functions code \citep[SIR;][]{cobo+toroiniesta1992}. We used a single magnetic component in 
the umbra including the light bridge, two magnetic components in the penumbra and one field-free and one 
magnetic component in quiet Sun regions, wherever the polarization degree exceeded a certain threshold. All 
parameters of the magnetic field (field strength $B$, inclination $\gamma$, azimuth $\chi$) and the 
LOS velocity were constant with height. The stray light contribution, the relative fill factor 
for multiple components and the macroturbulent and microtubulent velocities were also used as free fit 
parameters \citep[see, e.g.,][]{beckthesis2006,beck2008,beck+rezaei2009}. 

All spectroscopic data in \ion{Ca}{2} IR 854\,nm from either SPINOR or IBIS were inverted with the version 
of the CAlcium Inversion based on a Spectral ARchive \citep[CAISAR;][]{beck+etal2015,beck+etal2019a} code 
that assumes local thermodynamic equilibrium (LTE). 
For both the LTE and NLTE version, the CAISAR code can reproduce observed spectra to a level of a few percent 
of $I_c$ \citep{2014ApJ...788..183B,beck+etal2015,beck+etal2019a}. We converted the rms 
difference of observed and best-fit spectra to the corresponding rms variation in temperatures at different 
optical depth levels as an error estimate (calculation courtesy of J. Jenkins; details will be published 
elsewhere). This yielded an rms temperature error of $1\sigma$ of about 20\,K for $\log \tau > -2$, an 
increase from 20 to 100\,K over the range $-2 > \log \tau > -5$ and $100-150$\,K for $-5 > \log\tau > -6$. 
The comparison of LTE and NLTE results of the same spectra showed a close agreement for $\log \tau > -3$ 
\citep{beck+etal2015} and a difference in absolute temperature values by a factor of about 2 at $\log \tau = -5$.
All NLTE temperatures given here were derived from the LTE results by a multiplication with the NLTE 
correction curve for the umbra as given in Figure 6 of \citet{beck+etal2015}.

The spectropolarimetric \ion{Ca}{2} IR data from 
SPINOR were subsequently inverted with the same simplified approach as in 
\citet{beck+choudhary2019} assuming an exponential decay of the magnetic field with optical depth in the 
form of $B(\log\,\tau) = B_0 \cdot \exp(-\log\,\tau / \Delta \tau)$ with the free parameters $B_0$ and 
$\Delta \tau$ while keeping the temperature stratification fixed to the output of the prior inversion step.

\begin{figure*}[!ht]
\centerline{
\includegraphics[angle=90,width = 1.03\textwidth]{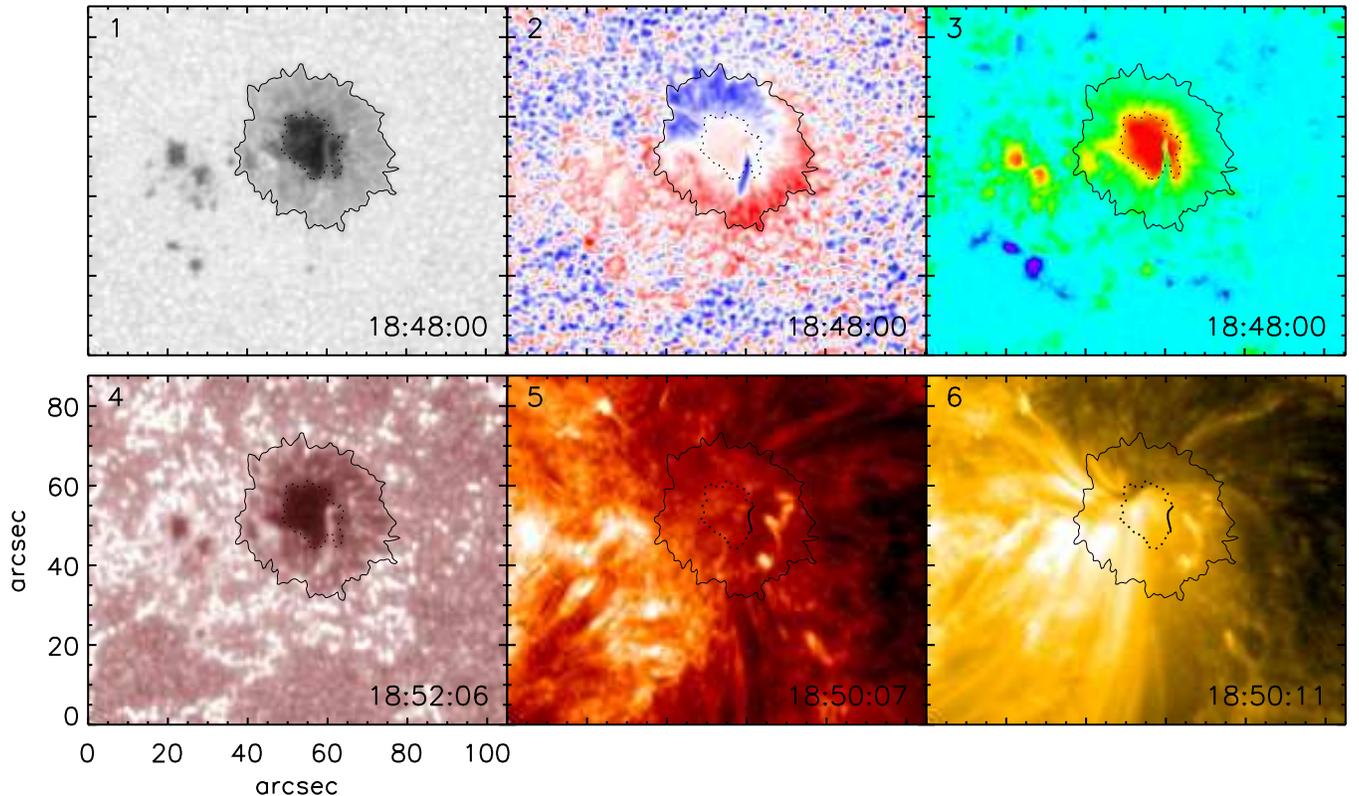}
}
\vspace{-45pt}
\caption{Leading sunspot in NOAA AR 12002 on 2014 March 13. Top row, from left to right : HMI continuum 
intensity, LOS velocity and the vertical component of the magnetic field.  Bottom row, from left to right : 
AIA 1700\,\AA, 304\,\AA, and 171\,\AA~channels. An animation of the temporal 
evolution of the sunspot from March 12 UT 23:00 until March 14 UT 05:40 is available in the online material.}
\label{fig01r}
\end{figure*}
   
\section{Results}
\label{res}

\subsection{General Overview of the LB and its Surroundings}
\label{overview}
NOAA AR 12002 transited the solar disk from 2014 March 08 to 2014 March 19. During the early phase of the AR's 
emergence between March 10 to March 12, a large sunspot in the leading polarity developed from several umbrae 
within a common penumbra, while several LBs were present in both the leading and follower spots. At its fully 
developed state, the leading polarity of the AR comprised two sunspots, the smallest of which developed and 
fragmented over 48\,hrs between March 12 to March 13. Our region of interest is a LB that 
appeared in the larger spot of the leading polarity of the AR, when the latter was fully developed and did 
not exhibit a significant change in its area. The LB in the leading sunspot was transient and there 
were no major flares or CMEs during its lifetime.

Figure \ref{fig01r} shows the leading sunspot of positive polarity in NOAA AR 12002 on 2014 March 13 
at 18:48 UT. There are a couple of pores ($x$,$y$: 25\arcsec, 45\arcsec) 
of the same polarity as the sunspot to the east of it. The 
penumbral region closest to these pores appears perturbed. 
To the south-east of the leading sunspot is a smaller set of pores of opposite polarity
($x$, $y$: 22\arcsec, 25\arcsec)
that are further out from the visible sunspot boundary (panel 3). The sunspot comprises a LB, about 
10\arcsec~in length, that runs along the north-south direction with its axial span very close 
to the western umbra-penumbra boundary encompassing the smaller umbral core of about 13\,\% of 
the total umbral area. The LB shows photospheric blue-shifts of about 0.5--0.7\,km\,s$^{-1}$ 
from its northern end to the mid-penumbra in the southern
section of the sunspot (panel 2) while the outer penumbra shows red-shifts of about 0.7\,km\,s$^{-1}$ arising from 
the photospheric Evershed flow \citep{1909MNRAS..69..454E}. 
In addition a small-part of the northern end of the LB harbors very weak red-shifts of about 0.2\,km\,s$^{-1}$
($x$, $y$: 60\arcsec, 54\arcsec) at the location where the LB meets the umbra-penumbra boundary. Panel 3 shows that the
vertical component of the magnetic field in the LB is greatly reduced in comparison to the
adjacent umbra where the values are about 1700\,G. In the mid-section of the LB the vertical
component is about 450\,G at its center and about 900\,G at the edges. Further down the LB, 
the values are even smaller, ranging between 115\,G to 260\,G, while those in the adjacent 
penumbra are about 650\,G.

The lower panels of Figure \ref{fig01r} show the LB in different AIA channels. The 1700\,\AA~channel 
from the upper photosphere (panel 4) indicates that the LB is brighter than the adjacent penumbra having 
a higher contrast than in the HMI continuum intensity. The He {\sc ii} 304\,\AA~image 
from the upper chromosphere (panel 5) shows that there are several surges which start from the right/western 
edge of the LB ($x$, $y$: 61\arcsec, 50\arcsec) and extend close to the sunspot boundary. These surges are 
seen all along the length of the LB and are characterized by dark filamentary structures along their length 
terminating in intense bright blobs.  The bright blobs are are also visible in the AIA 171\,\AA~image 
although the filamentary structure of the surges are more diffuse. 

An animation of the temporal evolution from March 12 UT 23:00 until March 14 UT 05:40 in the same quantities as 
in Figure \ref{fig01r} is provided in the online material. The relation between the magnetic flux emergence 
and the LB, the evolution of bright grains and surges and the sunspot rotation are more obvious in the 
animation than in the stills used for the subsequent figures.

\subsection{LB Formation}
\label{form}
In this section, we analyze the events that led to the formation of the LB in the leading sunspot.

\begin{figure}[!ht]
\centerline{
\includegraphics[angle=0,width = \columnwidth]{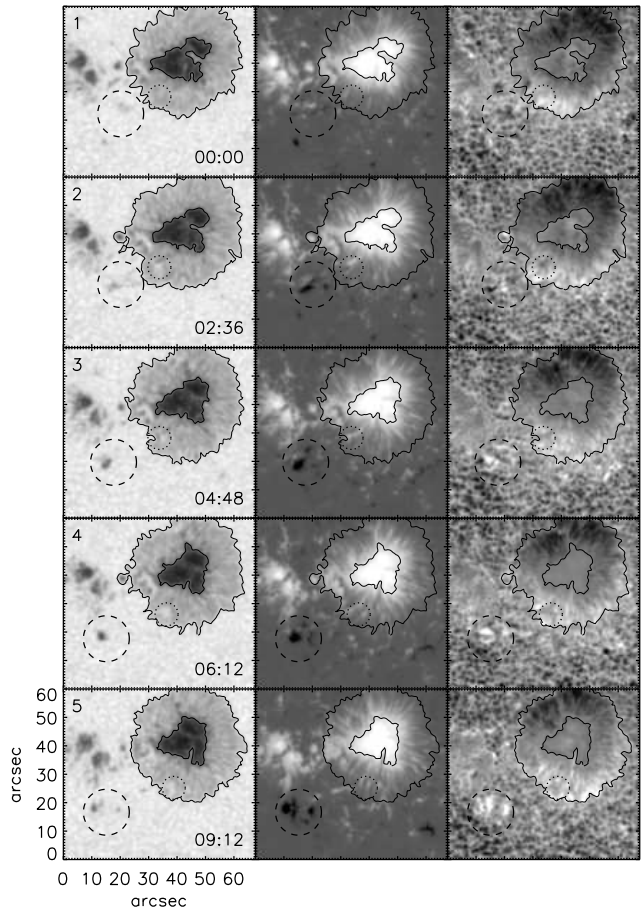}
}
\vspace{5pt}
\caption{Emergence of pore in the QS just outside the leading sunspot on 2014 March 13 (large dashed circle). 
The left, middle, and right columns correspond to the continuum intensity, $B_z$, and LOS velocity
respectively. The smaller dotted circle indicates changes in the sunspot penumbra associated with the pore's
emergence.}
\label{fig02r}
\end{figure}

\begin{figure}[!ht]
\centerline{
\includegraphics[angle=90,width = \columnwidth]{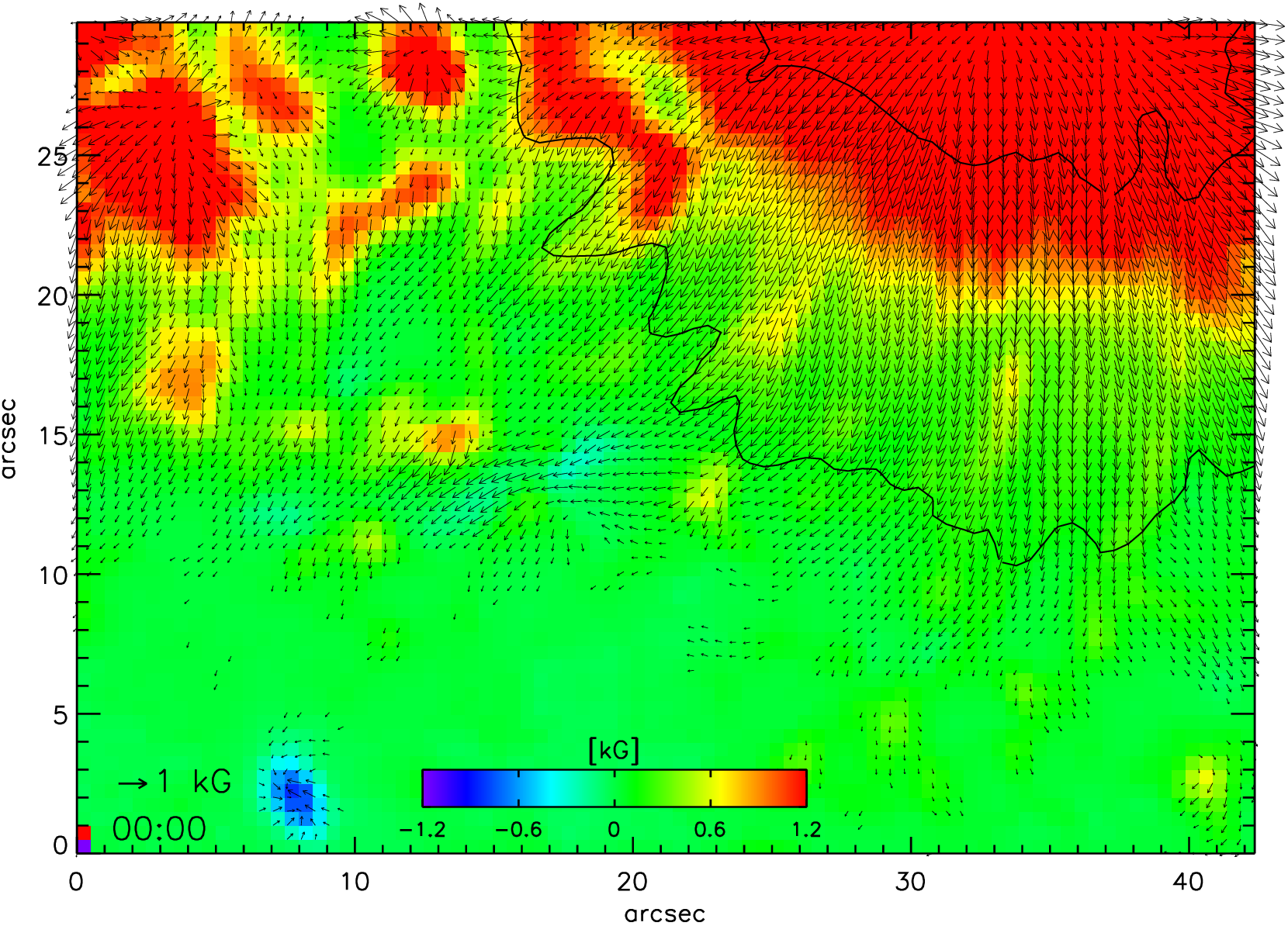}
}
\centerline{
\includegraphics[angle=90,width = \columnwidth]{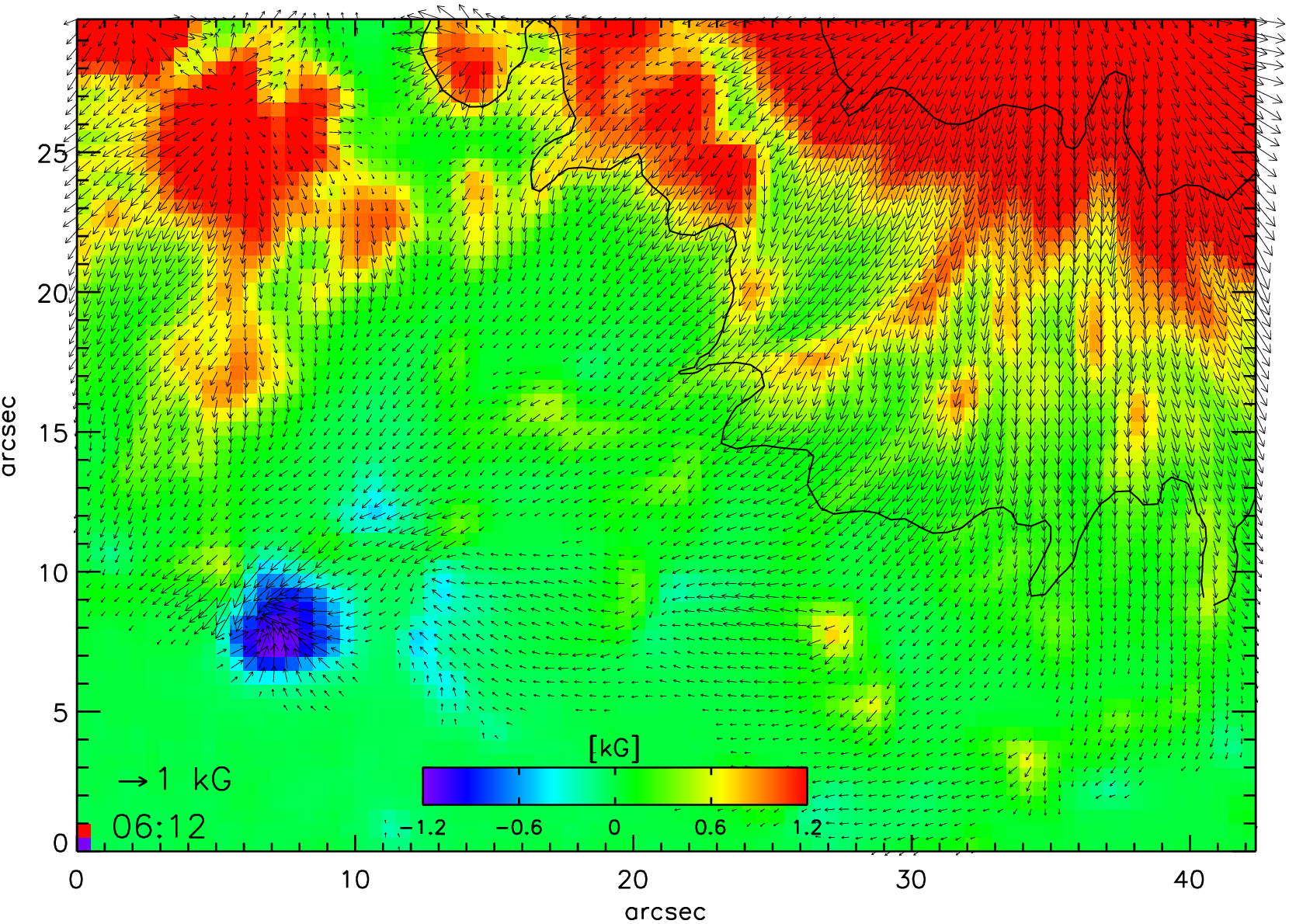}
}
\vspace{-5pt}
\caption{Horizontal magnetic field overlaid on the vertical component before and 
after the emergence of the pore at 00:00 UT (top) and 06:12 UT (bottom), respectively on 2014 March 13.
Arrows have been drawn for every pixel, where the field strength exceeds 200\,G.  
An example arrow corresponding to a field strength of 1\,kG is shown in the lower left corner 
above the timestamp.}
\label{fig03r}
\end{figure}

\begin{figure*}[!ht]
\centerline{
\hspace{10pt}
\includegraphics[angle=90,width = 0.94\textwidth]{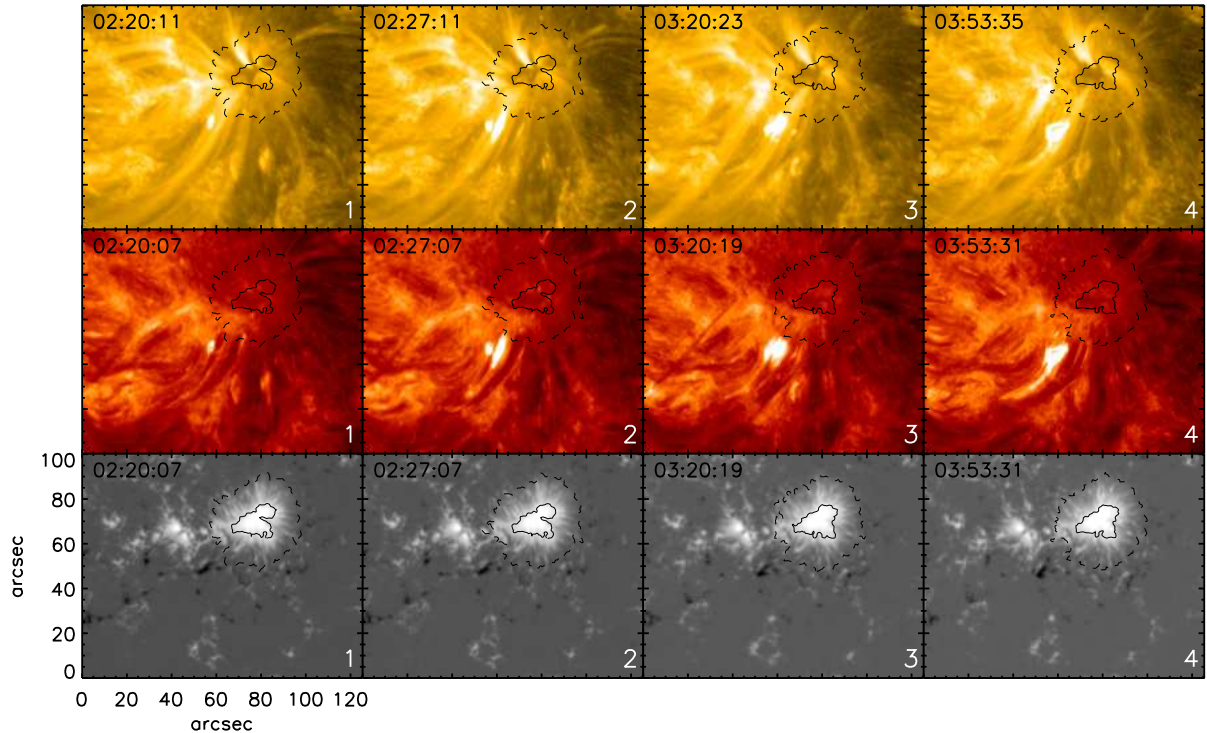}
}
\vspace{-70pt}
\caption{Response of the chromsophere and transition region to the emergence of the pore on 2014 
March 13. Top, middle, and bottom panels correspond to AIA 171\,\AA~and 304\,\AA~channels, and the 
HMI LOS magnetogram, respectively. The solid and dashed contours correspond to the umbra-penumbra 
and penumbra-quiet Sun boundary, respectively, of the HMI continuum intensity.}
\label{fig04r}
\end{figure*}

\subsubsection{Emergence of Pore}
\label{pore}
The top panels of Figure \ref{fig02r} show the leading sunspot at the beginning of 2014 March 13. The umbra
exhibits several bright intrusions at this point of time, specifically in the central-western part. 
The large dashed circle ($x$, $y$: 22\arcsec, 25\arcsec) in the panels marks 
a region just outside the spot boundary where a diffuse dark patch is seen that coincides with 
weak, mixed polarity fragments. The dotted circle ($x$, $y$: 33\arcsec, 28\arcsec) 
indicates a section of the outer penumbra that is involved in the magnetic flux emergence. Panel 2 of 
the figure shows the emergence of a negative polarity patch in the quiet Sun (QS), which is linearly 
extended in the direction of the sunspot as seen in the B$_z$ image. The continuum intensity image also 
shows dark features starting from the sunspot boundary and reaching up to the opposite polarity patch 
about 6\arcsec~away. In addition, the penumbra in the dotted circle exhibits a visible brightening 
below a filamentary structure (continuum intensity) that is oriented in the direction of the emergent 
flux in the QS. The LOS velocity image in panel 2 indicates weak blue-shifts 
of about 0.17\,km\,s$^{-1}$ at the location of this brightening. 

The newly emerged patch of opposite polarity
in the QS comprises red-shifts of about 0.75\,km\,s$^{-1}$ on the side closer to the spot and blue-shifts of 
about 0.55\,km\,s$^{-1}$ on the side furthest from the spot. Nearly 2\,hr later (panel 3) 
the opposite polarity patch has developed into a pore that is bigger, more circular, and has traversed 
further from the sunspot boundary. The penumbral sector closest to the emergent pore also exhibits a 
well-defined filamentary channel. We estimate the apparent horizontal speed of the pore, as it recedes 
from the sunspot, to be about 0.4\,km\,s$^{-1}$ from the time of its emergence at 04:00 UT . Panels 3 and 
4 of the continuum intensity also show 
that the conspicuous brightening in the outer penumbra progresses about 3\arcsec~to the west, while the 
region between the sunspot boundary and the pore is filled with small-scale mixed polarity fragments.
 
The pore exhibits strong red-shifts of about 1.4\,km\,s$^{-1}$, while the region in the penumbra marked with
the dotted circle is weakly red-shifted to about 0.1\,km\,s$^{-1}$ and clearly stands out because the surrounding
penumbra is dominated by red-shifts greater than 0.5\,km\,s$^{-1}$.

Following the data gap between 06:12--09:12 UT, we find additional negative polarity flux to have emerged 
close to the pore (Panel 5) which are associated with red-shifts of about 1.0\,km\,s$^{-1}$. On the other 
hand, the bright patch in the outer penumbra is associated with a kink in the filamentary penumbra located 
at the sunspot boundary. We observe that over a duration of about 9\,hr, starting at 00:00 UT on 2014 March 13,
the flux of the pore increases from -1.7$\times 10^{18}$\,Mx to -1.4$\times 10^{20}$\,Mx and the pore moves nearly
15\arcsec~away from its point of emergence. The sunspot umbra also exhibits fewer bright intrusions as 
seen in Panel 1 with only one set of penumbral filaments extending into the umbra. 

Figure \ref{fig03r} shows the horizontal magnetic field overlaid on the vertical component before and 
after the emergence of the pore at 00:00 UT and 06:12 UT, respectively. The vertical
component in the penumbral filamentary structure, aligned roughly in the direction of the pore at 
06:12 UT, has increased on an average by about 175\,G, while the horizontal magnetic field has reduced 
by about 180\,G. In addition to the pore, the transverse magnetic field indicates smaller loops between 
opposite polarity fragments, just west and north-west of the pore. While the filamentary penumbral structure
appears to be aligned in the direction of the pore, the difference between its orientation with the 
transverse magnetic field varies from about 15$^\circ$ in the mid-penumbra to about 40$^\circ$ in the outer
penumbra.

The emergence of the pore is associated with several transient events in the chromosphere and transition
region as shown in Figure \ref{fig04r}. Jet-like ejections are seen to emanate close to the emergent pore 
($x$, $y$: 60\arcsec, 50\arcsec) as well as just outside the sunspot boundary. AIA 304\,\AA~and 171\,\AA~images 
indicate that the jets occur beneath large-scale structures connecting the leading sunspot to the following 
polarity of the AR. The jets are possibly associated with flux cancellation in the vicinity of the emergent 
pore involving small-scale parasitic magnetic patches as they encounter the moving magnetic features 
streaming from the leading sunspot. Between 02:00 UT to 06:00 UT we detect six such jets with lifetimes 
of about 4--5 minutes, while some of the jets, such as those shown in panel 4, last for about 20\,min.

\begin{figure*}[!ht]
\centerline{
\includegraphics[angle=0,width = \textwidth]{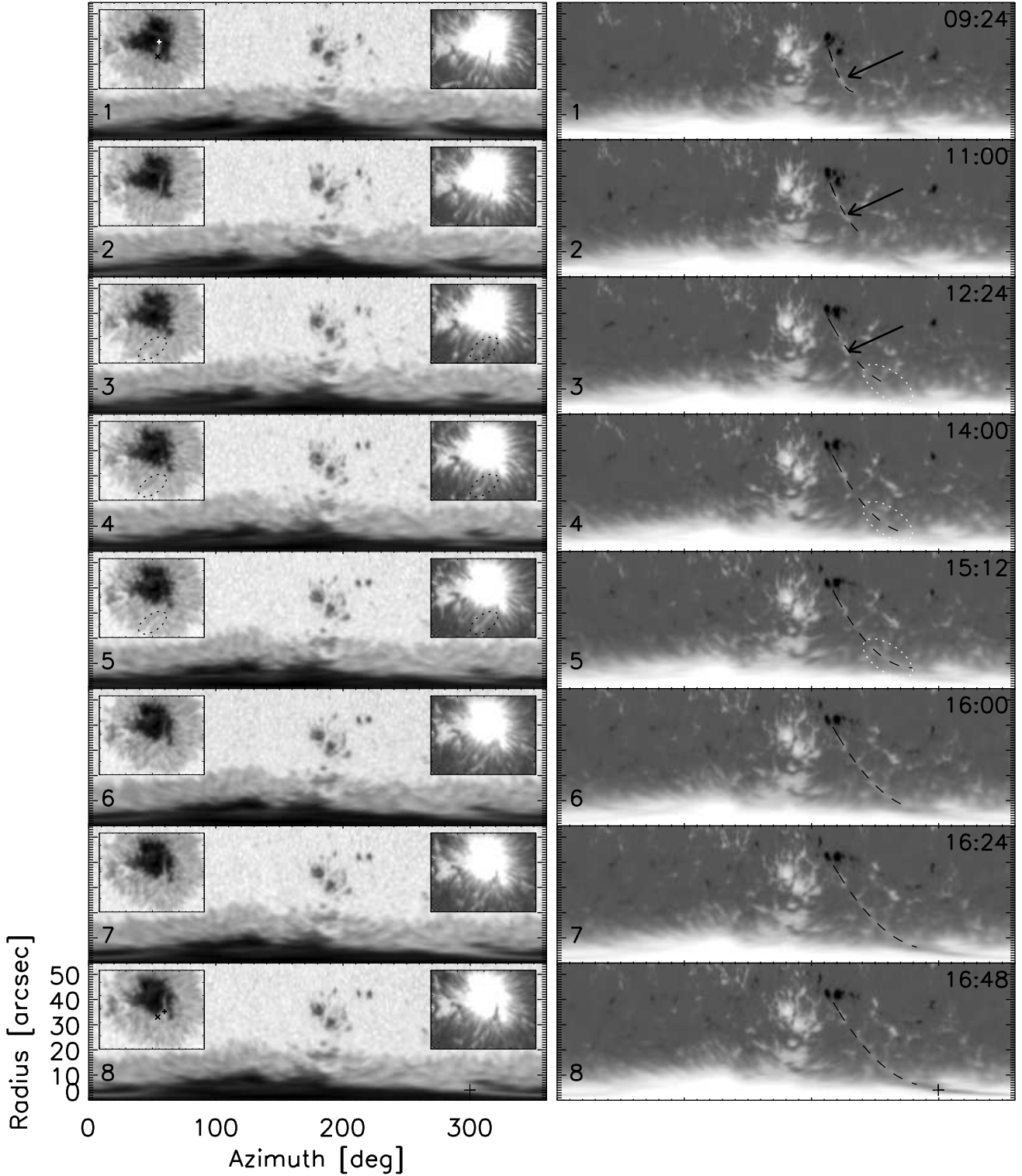}
}
\vspace{-130pt}
\caption{Formation of a light bridge in the leading sunspot on 2014 March 13. Time sequence of continuum 
intensity and the vertical component of the magnetic field in polar coordinates. The white plus symbol 
shown in the inset of panel 1 is the centre about which the polar transformation was carried out. The 
arrows in the right column indicate motion of positive polarity flux from the edge of the sunspot 
towards the pore of opposite polarity. The dotted ellipses in the right column indicate the development 
of a magnetic channel 
in the penumbra prior to the light bridge formation. The plus symbols in the bottom panels show the base 
of the light bridge in the image as well as its polar counterpart. The black cross symbol in panels 1 and 
8 indicate the location where the previous filamentary intrusion in the umbra existed. The black dashed 
line in the right panels indicates the magnetic association of the pore with the newly formed LB. The 
change from an almost straight (panel 1) to a strongly curved line (panel 8) results from the rotation 
of the sunspot.}
\label{fig05r}
\end{figure*}

\subsubsection{Precursor to Light Bridge Formation}
\label{intrude}
As shown in panel 5 of Figure \ref{fig02r}, a single set of penumbral filaments are seen to extend into the 
umbral core of the leading sunspot at a time when the pore in the QS has fully developed. Figure \ref{fig05r}
further shows the evolution of the leading sunspot using a polar transformation of the Cartesian image shown 
in the top left corner of the panel. The figure shows that the single set of penumbral filaments at 
09:12 UT on 2014 March 13 eventually breaks up by 12:24 UT (panel 3), which involves the filament to detach 
as several bright blobs or umbral dots that move further into the umbra. The detachment occurs at the northern end 
of the intrusion and progresses southwards towards the umbra-penumbra boundary with time. In addition, 
the kinked penumbral structure near the sunspot boundary, described in the previous section, fragments into
smaller patches that move further outwards closer to the pore (arrows in panels 1 to 3). Panels 3, 4, and 5 
show that, while the umbra is devoid of any intrusions from the penumbra, an elongated channel develops in 
the penumbra extending from the sunspot boundary to the inner penumbra (ellipse in right panels). Panels 6 
and 7 indicate this channel moving further into the umbra-penumbra boundary while simultaneously a bright
intrusion appears in the umbra. This intrusion develops rapidly over a duration of about 1\,hr starting at
15:48 UT (panel 8) while its base coincides with a relatively broad penumbral section (plus symbol) 
comprising several filaments. The base of this intrusion is slightly displaced from 
the location of the previous intrusion (cross symbols in panels 1 and 8). The polar image 
of the vertical component of the magnetic field in panel 8 shows that the new umbral intrusion can be traced 
outwards along the elongated magnetic channel in the penumbra and up to the pore in 
the QS (dashed line in right panels). This connection between the pore and the intrusion evidently deviates 
from a strictly radial alignment. 
The new structure that formed at 16:48 UT, 3.5 hrs after the earlier intrusion, is more extended 
as well as long-lived, and will henceforth be referred to as the light bridge (LB) for the remainder of 
the article.

\begin{figure}[!ht]
\centerline{
\includegraphics[angle=0,width = \columnwidth]{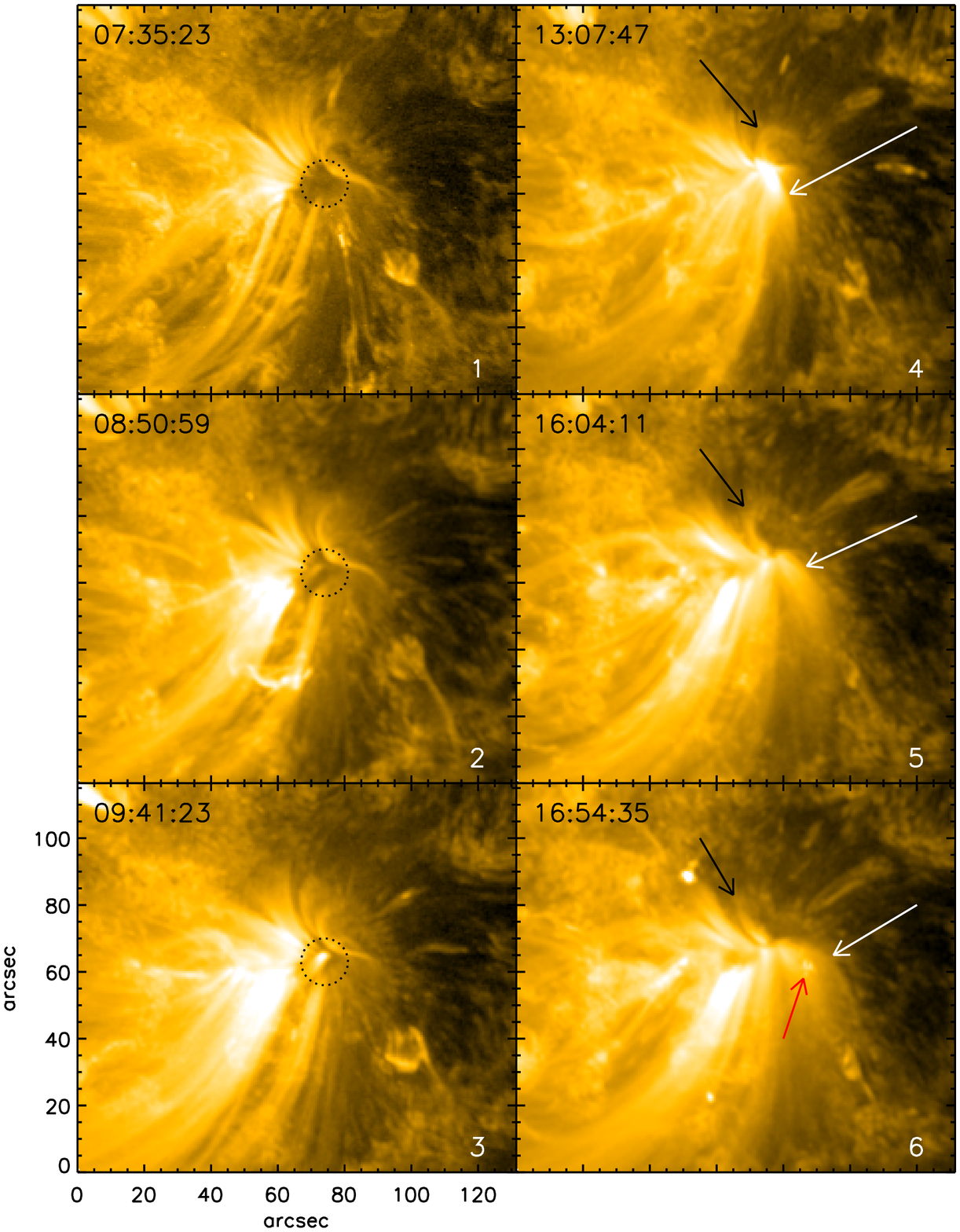}
}
\vspace{-30pt}
\caption{Evolution of large-scale loops in the leading sunspot on 2014 March 13. The dotted circle shows the 
onset of a bright loop whose one end is rooted in the umbra. The white and black arrows indicate the motion 
of large-scale loops in the counterclockwise direction. The red arrow corresponds to the surge from the light 
bridge.}
\label{fig06r}
\end{figure}

Figure \ref{fig06r} shows a sequence of AIA images in the 171\,\AA~channel leading up to the formation 
of the LB in the leading sunspot. The dotted circle in panel 1 shows a large part of the umbral 
core to be devoid of any large-scale structures. The half-'S' shaped loop at the top-right section of the 
dotted circle overlies the diffuse chain of umbral dots north of the penumbral intrusion (panel 1 of 
Figure \ref{fig05r}). A faint footpoint of a loop, rooted in the umbra, is seen around 08:50 UT (panel 2). 
This footpoint appears possibly as a response to a small flare that occurred in the AR around 08:35 UT. 
Panel 3 shows the footpoint getting brighter and the loop associated with the footpoint is seen to exhibit 
a counterclockwise rotation (white arrows; see also the animation in the online material). The umbra of 
the leading sunspot is now obscured by several loop footpoints as evident from the figure. The rotation 
of the loops is also observed in the northern sector of the sunspot (black arrows). We estimate that the 
loops rotate by about 40$^\circ$ over a duration of about 4\,hr. The red arrow in panel 6 shows the first 
signature of the surges emanating from the newly formed LB.

To summarize, the formation of the LB is preceded by the emergence of a pore nearly 17\,hr earlier just outside 
the sunspot boundary. The pore has an opposite polarity as the parent sunspot and drifts away from it at a speed 
of about 0.4\,km\,s$^{-1}$. Several jets occur in the chromosphere and transition region
in close proximity to the pore. The emergence of the pore is accompanied by an elongated magnetic channel in 
the penumbra which eventually reaches the umbra-penumbra boundary triggering the formation of the LB. The 
LB occurs close to the location where an umbral intrusion from the penumbra existed about 3.5\,hr earlier. 
The relation of the flux emergence to the subsequent formation of the LB is more obvious in the animation 
in the online material.

\begin{figure}[!ht]
\centerline{
\includegraphics[angle=0,width = \columnwidth]{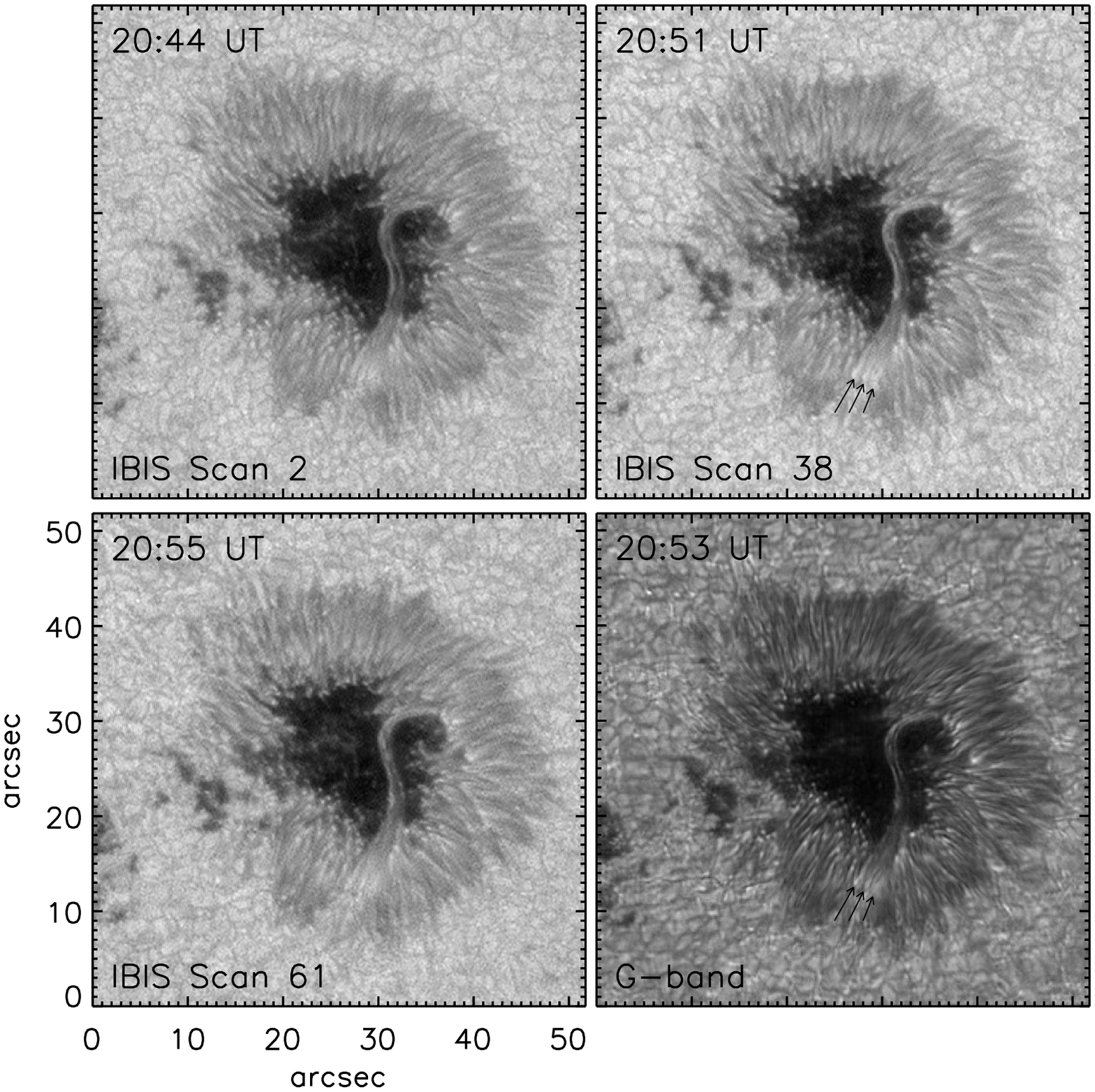}
}
\vspace{-90pt}
\centerline{
\includegraphics[angle=0,width = \columnwidth]{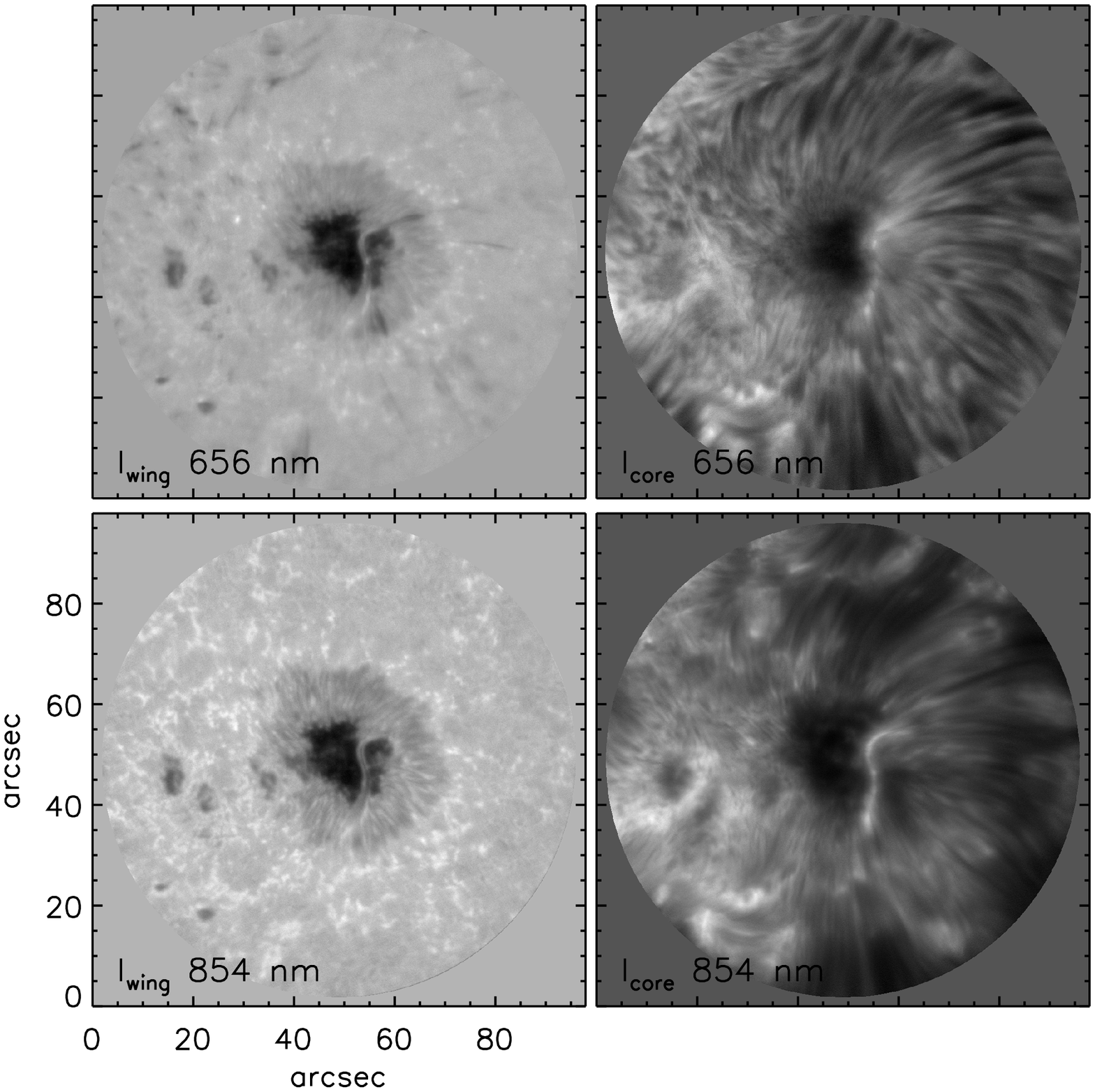}
}
\vspace{-90pt}
\caption{Top: speckle-reconstructed IBIS broad-band and G-band images taken from the DST on 2014 March 13. 
The three arrows in the right panels indicate bright grains in the outer penumbra with their corresponding 
filaments directed towards the umbra. Bottom: IBIS narrow-band filtergrams obtained in H$_\alpha$ (top) and 
\ion{Ca}{2} IR (bottom) at 20:51 UT on 2014 March 13 corresponding to spectral scan number 38. The left 
(right) column shows line-wing (line-core) images.} 
\label{fig06ra}
\end{figure}

\begin{figure}
\centerline{
\resizebox{7cm}{!}{\includegraphics[angle=0,width = \columnwidth]{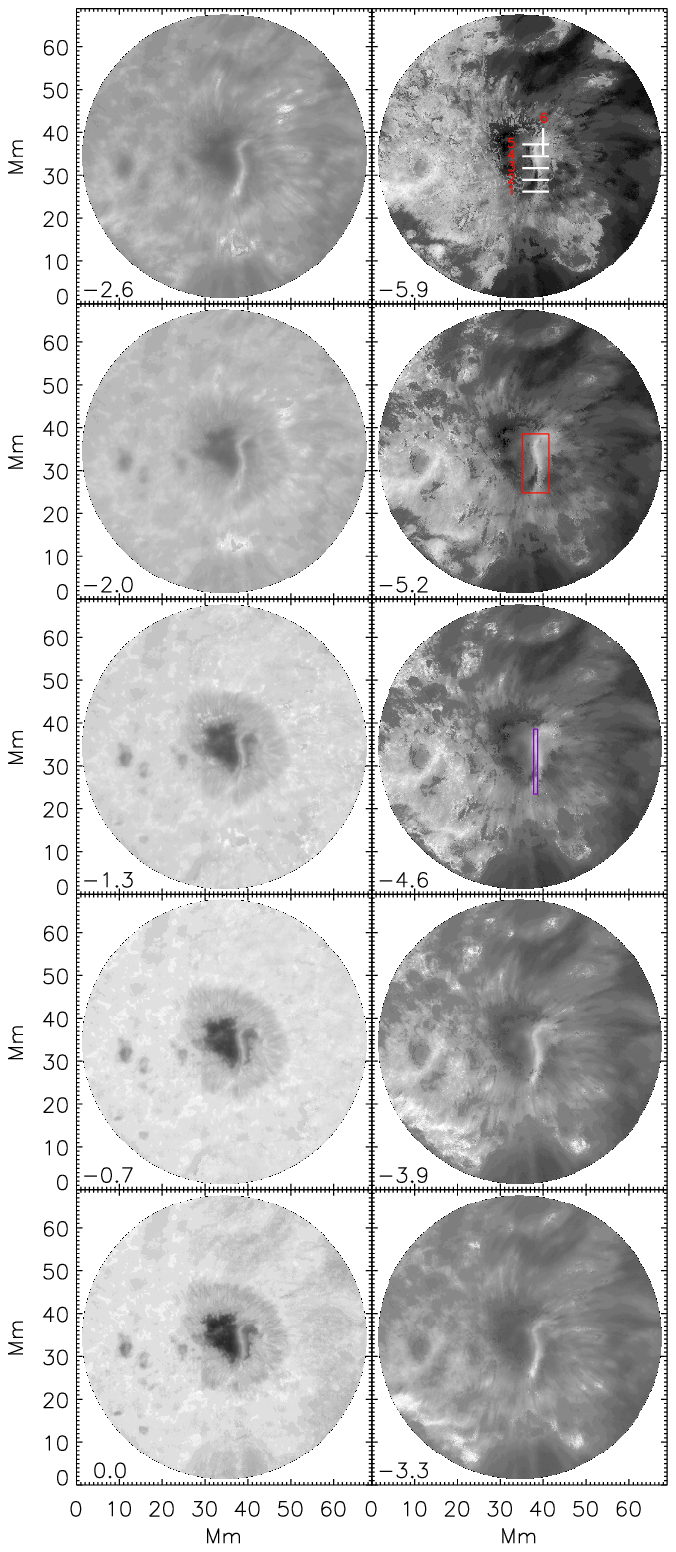}}
}
\vspace{5pt}
\centerline{
\hspace{20pt}
\resizebox{6cm}{!}{\includegraphics[angle=0,width = \columnwidth]{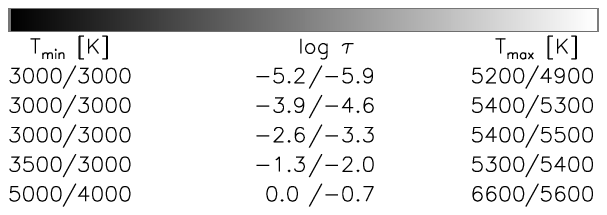}}
}
\caption{Temperature at different optical depths from the inversion of IBIS \ion{Ca}{2} IR spectra in 
spectral scan No.~2. The optical depth from $\log \tau = 0$ to $-5.9$ is given in the lower left corner 
of each panel while the temperature range at each optical depth is indicated below the color bar.
The purple/red rectangle in the third/fourth row of the right column marks the area 
of averaging in x or y used in Figures \ref{fig_cross} and \ref{fig_along}, while the locations of the 
cuts 1--6 across the LB of Figure\ref{fig_cross} are indicated in the upper right panel.}
\label{fig_ibis_overv}
\end{figure}

\begin{figure*}
\begin{minipage}{15.7cm}
\hspace{5pt}
\centerline{\resizebox{15.7cm}{!}{\includegraphics{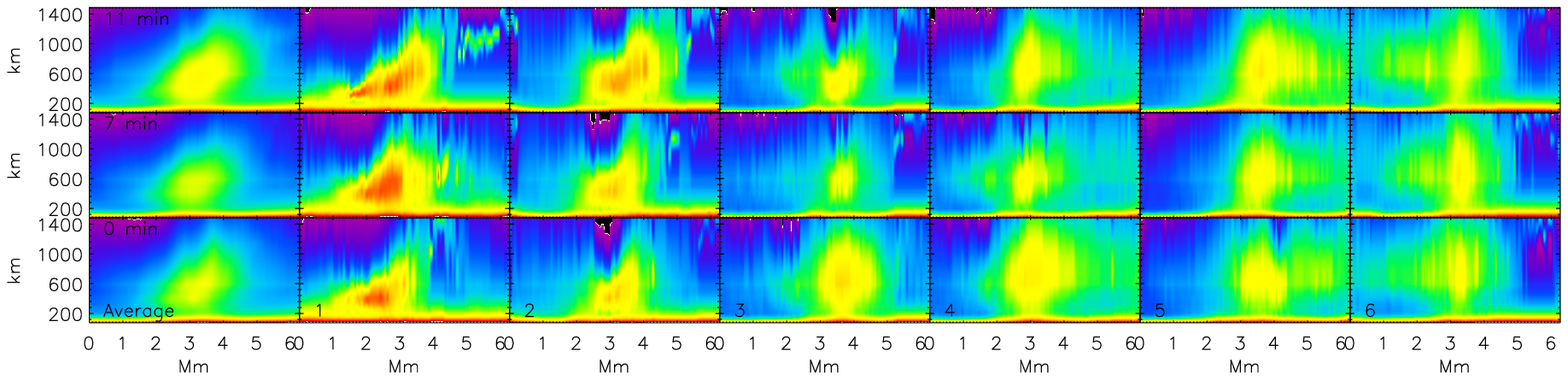}}}
\end{minipage}\hspace*{0.15cm}
\begin{minipage}{2.cm}
\vspace*{-0.5cm}
\resizebox{0.8cm}{!}{\includegraphics{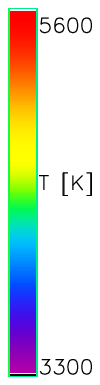}}
\end{minipage}
\vspace{-5pt}
\caption{Temperature from the inversion of IBIS \ion{Ca}{2} IR spectra on x-z cuts across the LB. Left to 
right: temperature of the LB cross-section averaged over the y-range indicated by the length of the red 
rectangle in Figure \ref{fig_ibis_overv} and temperatures at the cuts 1--6. Cuts are from East to West 
for Nos.~1-5 and South to North for No.~6. Bottom to top: results from spectral scans Nos.~2, 38, and 61 
at $t=0, 7$ and 11\,min. The color bar at the right-hand side shows the temperature range for all panels.}
\label{fig_cross}
\end{figure*}

\begin{figure*}
\begin{minipage}{15cm}
\hspace{10pt}
\centerline{\resizebox{15cm}{!}{\includegraphics{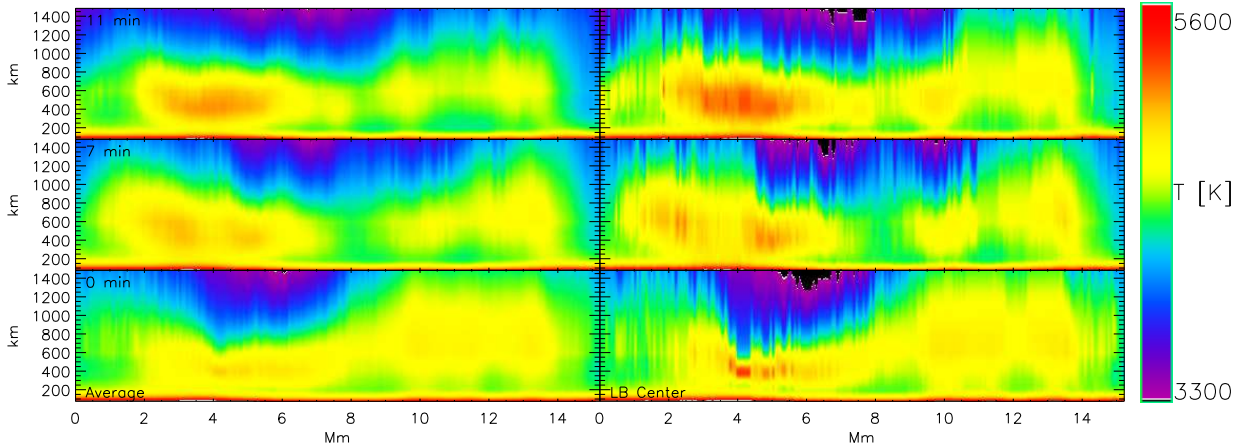}}}
\end{minipage}\hspace*{.3cm}
\begin{minipage}{2.cm}
\vspace*{-0.55cm}
\resizebox{1.3cm}{!}{\includegraphics{ibis_crosssection_zbar.eps}}
\end{minipage}
\vspace{0pt}
\caption{Temperature from the inversion of IBIS \ion{Ca}{2} IR spectra along the LB. Left column: 
temperatures along the LB averaged over the x-range indicated by the width of the purple rectangle 
in Figure \ref{fig_ibis_overv}. Right column: temperatures along the center of the LB. South is at 
0\,Mm. Bottom to top: results from spectral scans Nos.~2, 38, and 61 at $t=0, 7$ and 11\,min. The 
color bar at the right-hand side shows the temperature range for all panels.}
\label{fig_along}
\end{figure*}

\begin{figure}
\vspace{5pt}
\centerline{
\includegraphics[angle=0,width = \columnwidth]{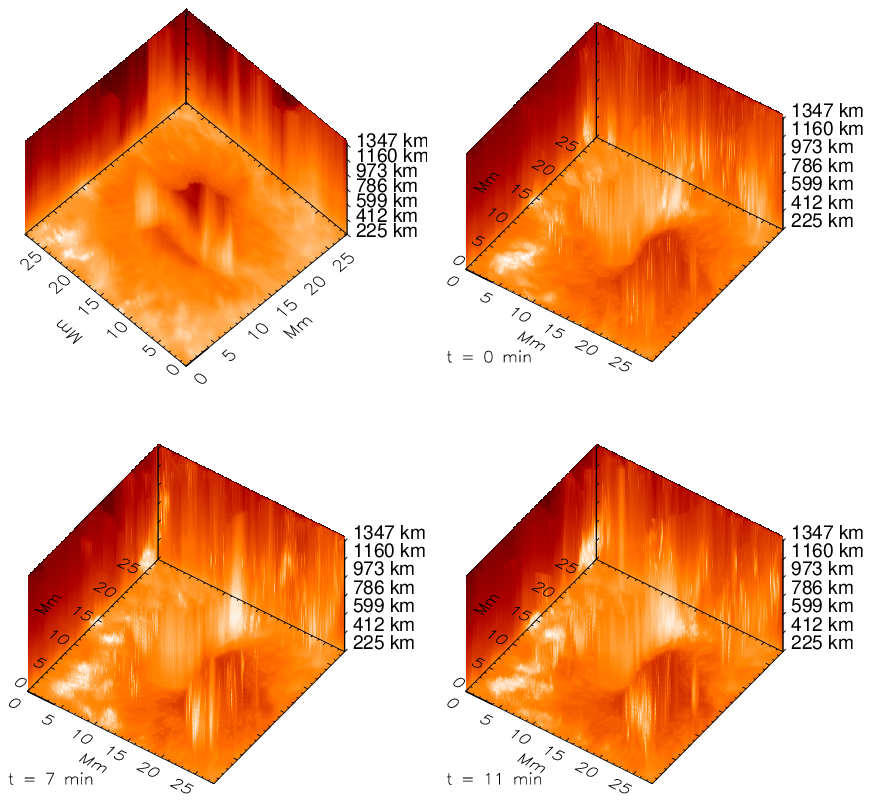}
}
\vspace{0pt}
\caption{3D temperature renderings of the LB. Left to right: SPINOR \ion{Ca}{2} IR thermal inversion 
results with a view across the smaller umbral core from South-West and IBIS inversion results of bursts 
Nos.~2, 38 and 61 at 0, 7 and 11\,min across the large umbral core from North-East.}
\label{3d_spinor}
\end{figure}

\subsection{LB Properties from DST Observations}
\label{dstprop}
In this section, we describe the thermal, kinematic, and magnetic structure as derived from the DST data 
from multiple instruments. These results illustrate a snapshot in the long-term evolution of the LB 
on 2014 March 13 at about 20:50 UT and provide a 3D perspective of the LB.

\subsubsection{LB Morphology: Shape in Photosphere and Chromosphere}
\label{dst_gband}
Figure \ref{fig06ra} shows the leading sunspot observed from the instrument suite at DST. The 
speckle-reconstructed images (top four panels) indicate the southern base of the LB at the inner 
penumbral boundary to be devoid of any penumbral features such as dark cores or penumbral grains 
although the average intensity is similar to that of the neighboring penumbra. The southern end of 
the LB is characterized by a couple of relatively bright filamentary structures (arrows in right 
panels). They resemble ''reverse'' penumbral filaments, with bright grains on the outer end of the 
penumbra, and then a photospheric penumbral filament going towards the umbra in the opposite direction 
to the regular sunspot structure. The northern part of the LB at this point in time has developed into 
an `S'--shaped hook that terminates at the inner penumbral boundary to the West. The LB is seen as 
extension of the penumbra running across the umbral core of the leading sunspot. There are no indications 
of dark lanes perpendicular to the long axis of the LB. There are no discernible changes over a period of 
11\,min spanned by the IBIS Broadband speckle reconstructions.

The bottom panels of Figure \ref{fig06ra} show the spectral scans taken with IBIS in H$_\alpha$ and 
\ion{Ca}{2} IR. In the wing of the Ca 854.2\,nm line the LB exhibits a central dark lane along its axis 
in the middle part of the LB, as also seen in the speckle-reconstructed images, while the line-core image 
shows the entire LB from its southern end to the `S'--shaped hook in the north to have an enhanced intensity 
compared to the rest of the sunspot. The `S'-shaped hook is oriented at nearly 90$^\circ$ to the axis of the 
LB and is seemingly continued by a dark fibril in the H$\alpha$ line-core image. The curvature in the southern 
end of the LB, however, is mostly radial. A similar feature is seen in the H$_\alpha$ line-core image although 
the brightness contrast is reduced. The blue line wing in H$_\alpha$ shows signatures of the surges on the LB 
(Figure \ref{fig01r}), at its southern and northern end. In addition there are also intense bright patches on 
the LB.

\subsubsection{Thermal Structure of LB}
\label{dst_thermo}
Figure \ref{fig_ibis_overv} shows the \ion{Ca}{2} IR LTE inversion results of the IBIS spectral scan 
No.~2 at 20:44 UT, on March 13. The LB stands out at all heights, both in the photosphere and chromosphere 
between $\log\tau=-2$ to $\log\tau=-5$, and its shape is retained in height as well. 
The figure also shows that the LB is hotter in the chromosphere than the rest of the sunspot. The 
enhancement in temperature relative to the umbra at $\log\tau = 0, -1, -2, -3, -4$, and $-5$ are 
452(475)\,K, 385(398)\,K, 402(406)\,K, 595(679)\,K, 774(1036)\,K, and 637(1356)\,K, respectively, where
the numbers in the parantheses correspond to the NLTE values. At $\log\tau=-3, -4,$ and $-5$, the maximum 
temperature in the LB is 4361(4978)\,K, 4390(5879)\,K, and 3985(8478)\,K, respectively. 
The maximal difference relative to the 
umbra is 600--800\,K and occurs at about $\log\tau=-3.6$. This enhancement in temperature along the LB is seen 
throughout the IBIS time sequence of 15\,min. The bright filamentary penumbra at the southern end of the LB is 
associated with an enhancement of 200\,K for $\log\tau\ge-2.5$ ($\sim 280$\,km), while it is cooler than the 
penumbra by about 300\,K at higher heights. From $\log\tau=-2.5$ to $\log\tau=-3.5$ the LB is hotter than all 
other regions in the FOV, including the quiet Sun, by 600--800\,K. 

Figures \ref{fig_cross} and \ref{fig_along} show 2D $x$--$z$ and $y$--$z$ plots of the temperature on cuts across 
and along the LB, respectively for the three IBIS spectral scans Nos.~2, 38, and 61 at $t= 0, 7$ and $11$\,min. 
These scans correspond to 
20:44 UT, 20:51 UT, and 20:55 UT on March 13, respectively. The lateral width of the LB cross-section is about
1--2\,Mm, while the LB extends to a height of about 1\,Mm. Cuts 1 and 2 which correspond to the southern end of the 
LB, reveal the thermal distribution across the LB to be quite complex, with the boundary enclosing the hot structure 
increasing in height from east to west. Cuts 3--6 show a more simpler, homogeneous thermal structure across 
the axis of the LB. Interestingly, these cuts also exhibit the loss of a continuous connection of the temperature 
enhancement to the photosphere, with the LB appearing nearly isolated at mid chromospheric heights. 

The highest 
temperature is usually found at the center of the structure, e.g., cut No.~2 (top row), cut No.~3 (bottom row), 
and cut No.~5 (top row). Proceeding from cuts 1 to 6, we find that the center of the temperature enhancement 
corresponding to the LB moves from the photosphere at about 400\,km to the lower chromosphere at 600\,km. 
Cuts 1 and 2 show that the LB is hotter in comparison to the remaining cuts at a height range of 300--700\,km.
In general the temperature distribution 
in the LB is highly inhomogeneous along and across its axis, as well as in height, which can also be visualized 
in Figure \ref{fig_along}. The maximum temperature along cuts 1, and 2 across the LB are about 5293\,K, and 
5075\,K, while the same for cuts 3--6 are about 4800\,K. The height at which the maximum temperature is observed 
on the LB is around 300--400\,km ($\sim\log\tau=-3$) for cuts 1 and 2 while the same for cuts 3--6 is at 
600--700\,km ($\sim\log\tau=-4$).

\begin{figure*}
\centerline{
\resizebox{16cm}{!}{\includegraphics{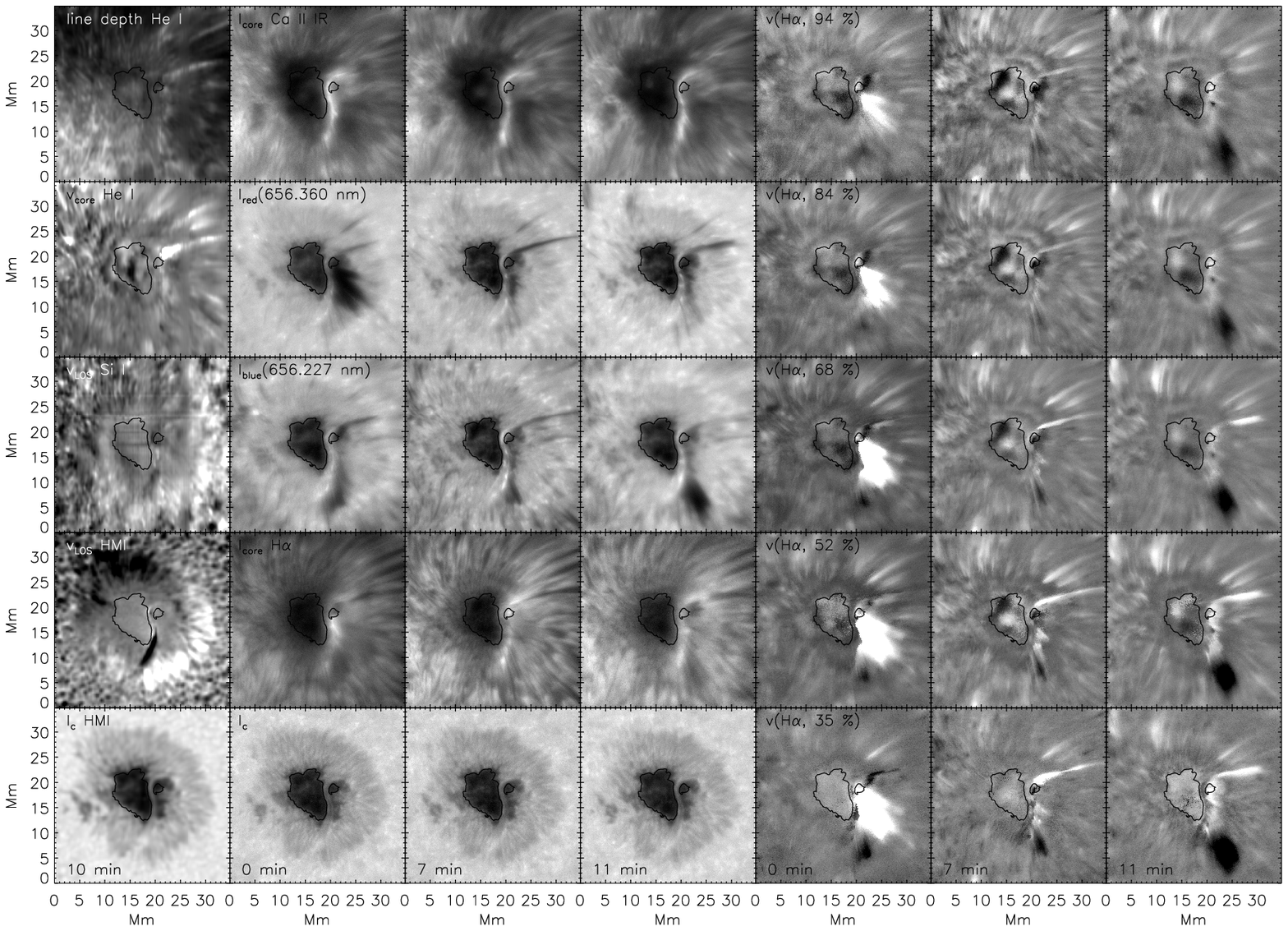}}
}
\vspace{-5pt}
\caption{Line-of-sight velocities and intensities in photospheric and chromospheric spectral lines from 
SPINOR and IBIS. Column 1, bottom to top: HMI continuum intensity, HMI LOS velocity, SPINOR \ion{Si}{1} 
LOS velocity, SPINOR \ion{He}{1} line core velocity, and SPINOR \ion{He}{1} line depth. Columns 2--4 bottom 
to top: IBIS continuum intensity, H$_\alpha$ line core intensity, H$_\alpha$ blue wing intensity, H$_\alpha$ 
red wing intensity, and IBIS \ion{Ca}{2} line core intensity. Columns 5--7, bottom to top:  bisector velocity 
of H$_\alpha$ at 35, 52, 68, 84 and 94\,\% line depth. The time step of each column is indicated in the lower 
left corner of each panel in the bottom row. Black contour lines outline the dark umbral cores.}
\label{fig_velo_ibis}
\end{figure*}

We find that the height variation of the temperature increases nearly monotonically along the length of the 
LB at $t=0$\,min. This changes at $t = 7,11$\,min where there is a bump close to the middle of the LB 
(Figure \ref{fig_along}). At all three instances of time, the temperature enhancement at the southern end 
of the LB, with respect to the umbra, is at a lower height with values of about 1100\,K at $\log\tau\sim-3.5$, 
while at the northern end, the values are about 640\,K. The highest temperature near this southern section is 
at a height of about 500\,km. The figure also indicates that the thermal enhancement at the southern end of the 
LB becomes elevated from $t=0$ to $t=7, 11$\,min between $x=2$--$6$\,Mm and extends over a height range 
of about 1\,Mm. A cold upper region at the middle of LB ($x=4$--$8$\,Mm) is located at about 600\,km at $t=0$\,min 
(right column, bottom panel), which moves up to 800\,km at $t=7$\,min, and reaches 900\,km at $t=11$\,min (right 
column, top). 

Figure \ref{3d_spinor} shows 3D renderings of temperature from the SPINOR and IBIS data. The southern end of LB 
is hotter than the northern end all the time. The `S'-shaped hook at the northern end is clearly seen as a 
continuation of higher temperature oriented at nearly 90$^\circ$ to the long axis of the LB (top left panel).
The structure of the LB stays similar over the 11\,min of DST observations with little differences between the IBIS 
and SPINOR  results apart from the higher spatial resolution of IBIS.

\begin{figure*}
\centerline{
\hspace{15pt}
\resizebox{16cm}{!}{\includegraphics{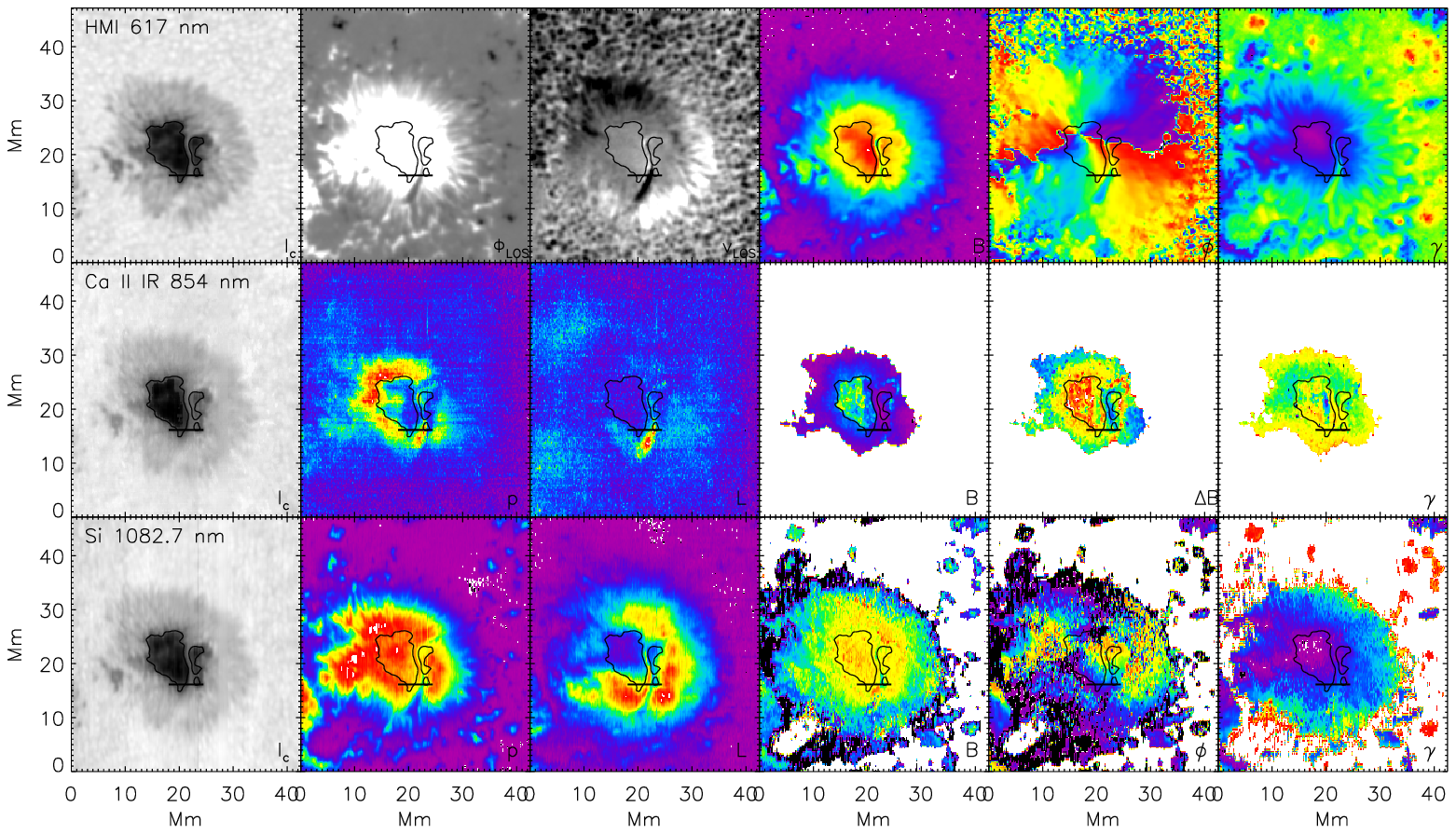}}
}
\vspace{-5pt}
\caption{Polarization signal and magnetic inversion results of photospheric and chromospheric spectral 
lines from HMI and SPINOR. Bottom row, left to right: continuum intensity $I_c$ at 1083\,nm, polarization 
degree $p$ and linear polarization $L$ of \ion{Si}{1} at 1082.7\,nm, magnetic field strength $B$, azimuth 
$\Phi$ and inclination $\gamma$ from the SIR inversion of \ion{Si}{1} at 1082.7\,nm. Middle row, 
left to right: $I_c$, $p$ and $L$ of \ion{Ca}{2} IR, $B$, decay constant $\Delta B$ and inclination 
$\gamma$ from the CAISAR inversion of \ion{Ca}{2} IR. Top row, left to right: $I_c$, LOS magnetic 
flux, LOS velocity, $B$, $\Phi$ and $\gamma$ from HMI. The horizontal black line ($x$, $y$: 20\,Mm, 16\,Mm) 
indicates the location of the cut shown in Figure \ref{fig_mag_cut}. Black contour lines outline the dark 
umbral cores.}
\label{fig_mag}
\end{figure*}

\subsubsection{Photospheric and Chromospheric LOS Velocities}
\label{dst_velo}
Figure \ref{fig_velo_ibis} shows persistent blue-shifts of about 0.31\,km\,s$^{-1}$ at the southern section 
of the LB as seen in the photospheric LOS velocity map derived from the \ion{Si}{1} line (panel 3, column 1). 
This is consistent with what is observed in the HMI data (panel 4, column 1), although the 
latter are more stronger and conspicuous in the red-shifted region of the limb-side penumbra. As seen earlier 
in Figure \ref{fig01r}, there are weak red-shifts at the northern end of the LB. These red-shifts are more 
diffuse in the \ion{Si}{1} line and are about 0.27\,km\,s$^{-1}$. The umbra is relatively at rest 
in the photosphere as indicated in both panels.  
Panel 2 of column 1 shows the velocity derived from the line core of the SPINOR \ion{He}{1} line. There are 
strong red-shifts of about 22\,km\,s$^{-1}$ that extend from a patch to the west of its northern end. 
These red-shifts are associated with the surges shown in Figure \ref{fig01r}. 
Red-shits in the rest of the sunspot are about 2--6\,km\,s$^{-1}$. In general, the chromospheric red-shifts 
follow the bright fibrils seen in the \ion{He}{1} line depth map (panel 1 column 1). The rest of the 
sunspot exhibits the regular and inverse EF in the photosphere and chromosphere, respectively.

The LB remains bright as seen in the \ion{Ca}{2} line-core image over a duration of 11\,min with several 
bright spikes directed to the west (panel 1 columns 2--4). Since the \ion{Ca}{2} spectra exhibit large 
emission and red shifts, the velocity values are unreliable.
Columns 5--7 of Figure \ref{fig_velo_ibis} shows the chromospheric LOS velocities derived from the 
IBIS H$_\alpha$ data. To put these velocities into context, H$_\alpha$ filtergrams in the line core, 
as well as in the blue and red wings are shown in the figure (panels 2--4, columns 2--4). The bisector 
velocities derived from the IBIS H$_\alpha$ data reveal rapid dynamic evolution of the surges in a span 
of 11\,min. A large section of the surges show red-shifts of about 20--25\,km\,s$^{-1}$ extending over 
an area of 31\,M$m^2$ at the 35\,\% level at min 0 (scan 02). The values are similar at 52\,\% and 68\,\% 
levels as well. However, close to the line core, at the 94\,\% level the red-shifts reduce to about 
10\,km\,s$^{-1}$ and occupy a smaller area of about 4\,M$m^2$. The red-shifts in this part of the LB 
also weaken significantly with time with blue-shifts of 11\,km\,s$^{-1}$ (at the 35\,\% level) which 
reduce to about 6\,km\,s$^{-1}$ as one approaches the line core. In contrast, the surges in the northern 
end of the LB along the bright fibril, initially show blue-shifts at lower line depths but change to 
red-shifts of about 6\,km\,s$^{-1}$ at $t=11$\,min.

\begin{figure}
\centerline{
\resizebox{8.8cm}{!}{\includegraphics{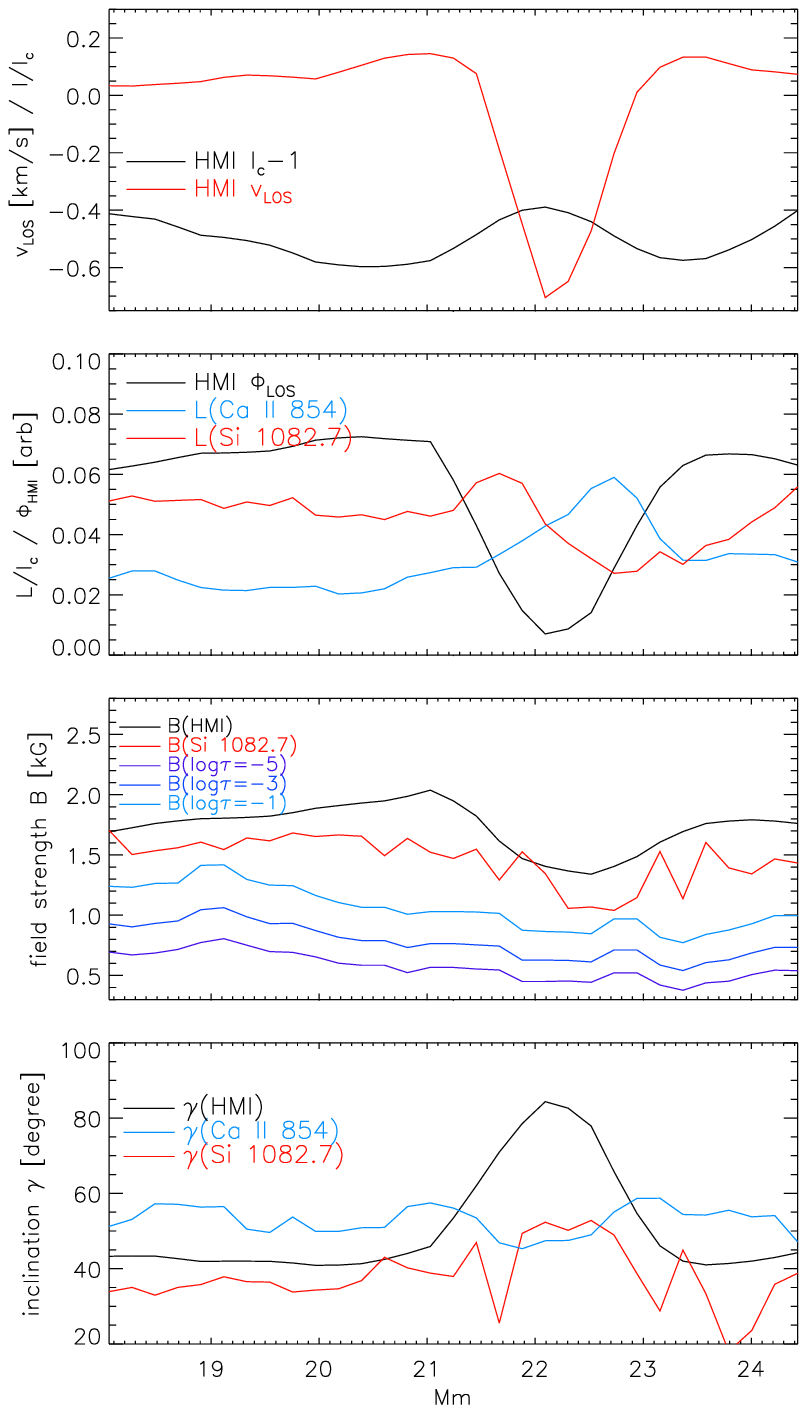}}
}
\vspace{-5pt}
\caption{Atmospheric properties along a cut across the LB. Top panel: continuum intensity$-1$ (black line) 
and LOS velocity (red line) from HMI. Second panel: LOS magnetic flux (black line) from HMI, linear polarization 
signal $L$ of \ion{Si}{1} at 1082.7\,nm (red line) and \ion{Ca}{2} IR (blue line). Third panel: magnetic field 
strength $B$ from HMI (black line), \ion{Si}{1} at 1082.7\,nm (red line) and at three optical depths of 
$\log \tau = -1, -3$ and $-5$ from \ion{Ca}{2} IR (blue to turquoise lines). Bottom panel: magnetic field 
inclination $\gamma$ from HMI (black line), \ion{Si}{1} at 1082.7\,nm (red line) and \ion{Ca}{2} IR (blue line). 
The cut is oriented East to West.}
\label{fig_mag_cut}
\end{figure}

\subsubsection{LB Magnetic Properties}
\label{dst_magneto}
Figure \ref{fig_mag} shows the magnetic field parameters derived from the \ion{Fe}{1} line of HMI as 
well as \ion{Si}{1} and \ion{Ca}{2} lines from SPINOR. In the photosphere, the LB is associated with 
highly inclined magnetic fields that are about 500\,G weaker than the adjacent umbra. However, it is 
at the location of the strong blue-shifts at the southern end of the LB, where the azimuth and inclination 
maps from HMI render the LB significantly distinct from the rest of the sunspot. In addition, the polarization 
degree and linear polarization derived from the \ion{Si}{1} line are highly reduced at that location. On 
the other hand, the above quantities are markedly enhanced when seen in the chromospheric \ion{Ca}{2} line.
 
Figure \ref{fig_mag_cut} shows the variation of the magnetic field over a cut across the blue-shifted patch 
in the southern end of the LB. The reduction in the magnetic flux seen in HMI is accompanied by an increase 
in the linear polarization in the \ion{Si}{1} line which is displaced to the east, while that in the 
\ion{Ca}{2} line is shifted to the west (top right panel). The field strength reduces with height in the 
LB, from 1300\,G in the HMI \ion{Fe}{1} line, to about 1000\,G in \ion{Si}{1}, and about 400\,G in the 
chromospheric \ion{Ca}{2} line. Figure \ref{fig_mag_cut} suggests that the magnetic field in the LB drops 
off much faster with height than adjacent regions of the sunspot. The same is indicated by the value 
of the decay constant $\Delta B$ in Figure \ref{fig_mag} that is smaller in the LB than its surroundings. 
While the magnetic field is highly inclined in the 
LB as seen from HMI, it becomes relatively vertical with height, reducing from about 85$^\circ$ to about 
45$^\circ$ in the \ion{Si}{1} line that samples the upper photosphere. However, the field inclination in 
the LB is relatively higher than its adjacent neighborhood at the formation height of the photospheric 
\ion{Fe}{1} and \ion{Si}{1} lines, as well as the chromospheric \ion{Ca}{2} line.


\begin{figure*}[!ht]
\centerline{
\vspace{20pt}
\includegraphics[angle=90,width = \textwidth]{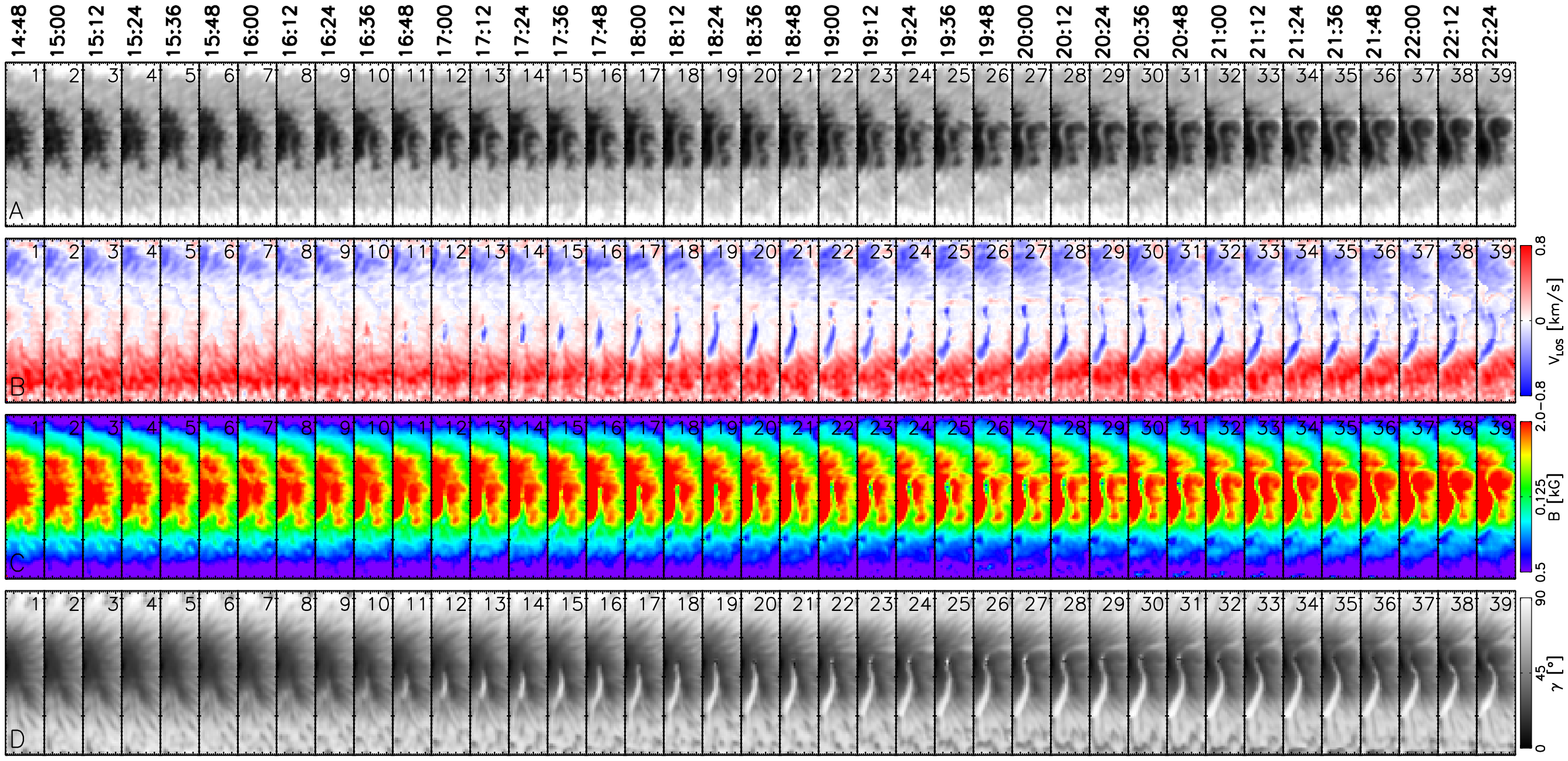}
}
\vspace{-170pt}
\centerline{
\includegraphics[angle=90,width = \textwidth]{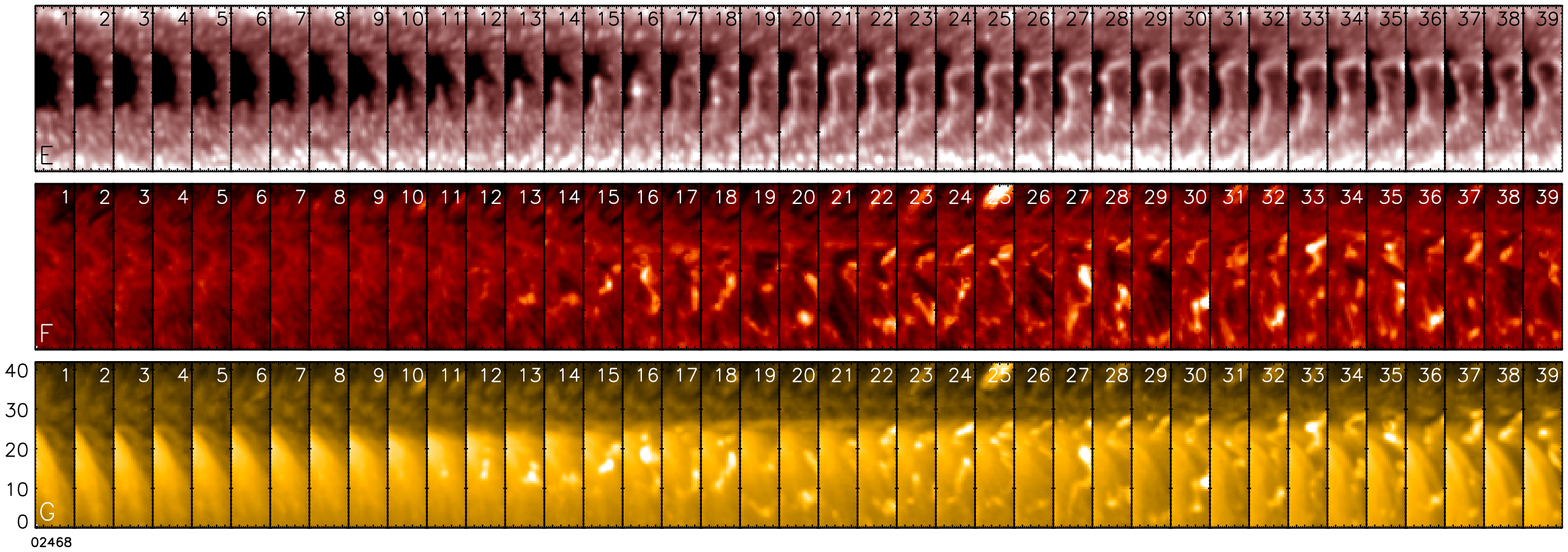}
}
\vspace{-175pt}
\caption{Temporal evolution of photospheric blue-shifts and relevant physical parameters in the light 
bridge from 2014 March 13--14. Rows A--D: HMI continuum intensity, LOS velocity, magnetic field strength, 
and magnetic field inclination. Rows E--G: AIA images in 1700\,\AA, 304\,\AA, and 171\,\AA.}
\label{fig07r}
\end{figure*}

\subsection{Long-term Temporal Evolution of the LB in SDO Data}
\label{evol}
As shown in the Section \ref{intrude}, the LB forms rapidly between 15:48--16:48 UT on 2014 March 13. 
This is indicated in Row A of Figure \ref{fig07r} which follows the temporal evolution of the LB
between 13 and 14 March. Panels 10 and 11 of Row B indicate weak photospheric blue-shifts of about 
0.13--0.32\,km\,s$^{-1}$ appearing at the broad base of the LB between 16:36 and 16:48 UT. 
These blue-shifts stretch along the LB northward into the umbra but also extend nearly 5\arcsec~southward 
from the umbra-penumbra boundary into the penumbra in about 2\,hrs. The maximum value of the blue-shifts 
at 18:48 UT is about 0.72\,m\,s$^{-1}$ (panel 21) while the axial length of the LB where the blue-shifts are
observed is about 10\arcsec. At this point of time, the northern end of the LB is anchored to the 
adjoining penumbra. The broad base in the penumbra, from where the LB began to form, appears as a 
smooth extension of the LB. Panel 21 in the continuum intensity shows that even at HMI's spatial 
resolution there are barb-like structures on the lower eastern edge of the LB ($x$, $y$: 7\arcsec, 18\arcsec). 

While the blue-shifts dominate
the LB, tiny patches of very weak red-shifts, of about 0.16\,km\,s$^{-1}$, are observed at a couple of locations 
on the LB. One of these red-shifted patches lies at the northern end of the LB close to the latter's anchorage
point at the umbra-penumbra boundary. This patch persists for at least 5\,hrs as seen in Figure \ref{fig08r}. 
In the animation and panels 10-20 one can see it as a persistent small red-shifted patch at the northern end of 
the LB that moves with its progression into the umbra. It could indicate the location of the other leg of the 
magnetic structure, similar to the pore with its red-shifts. The other red-shifted patches are comparatively 
more transient, lasting for about 30\,min, and are seen at the edge of the LB near its midsection 
(panels 22--25 of Figure \ref{fig07r}) and at the location of the barb described earlier.

\begin{figure*}[!ht]
\centerline{
\vspace{20pt}
\includegraphics[angle=90,width = \textwidth]{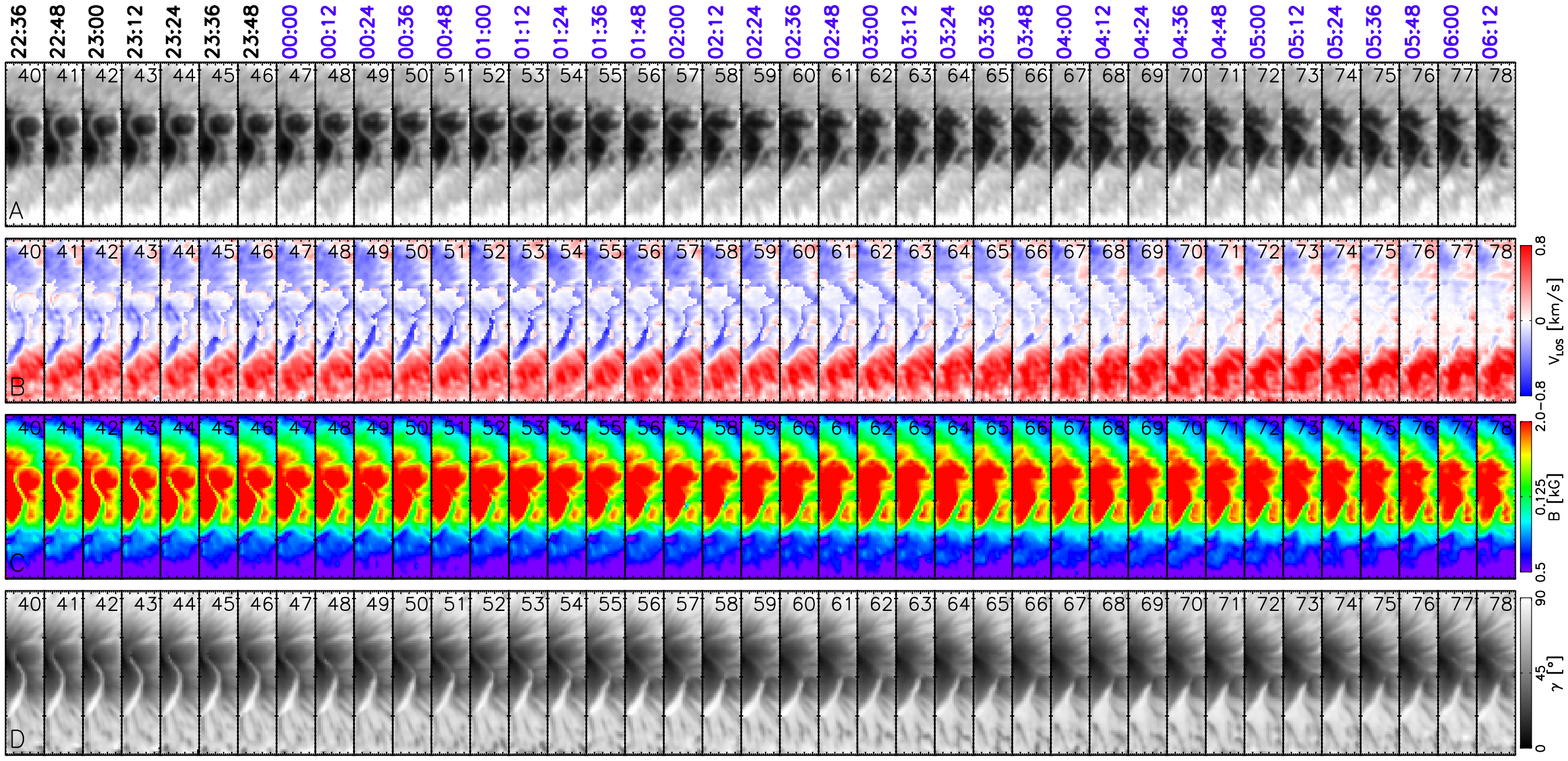}
}
\vspace{-170pt}
\centerline{
\includegraphics[angle=90,width = \textwidth]{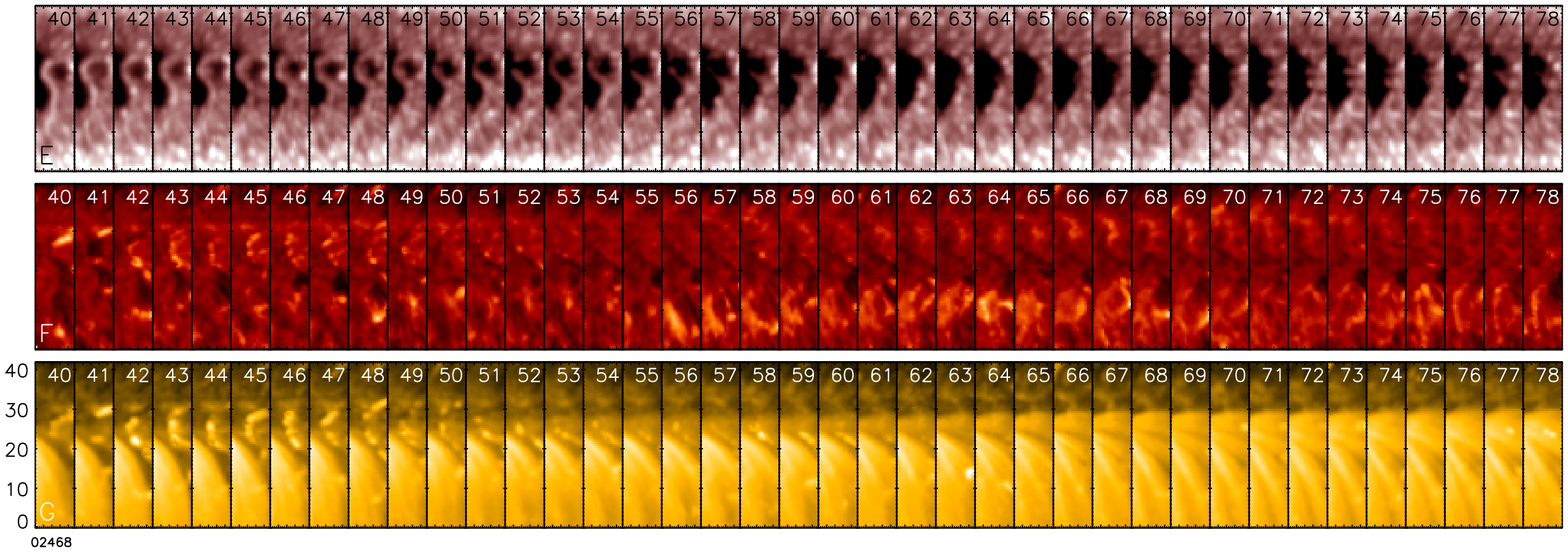}
}
\vspace{-175pt}
\caption{Continuation of Figure \ref{fig07r}. The purple color for the time on the top 
row signifies 2014 March 14.}
\label{fig08r}
\end{figure*}

Panels 26--39 show that the strongest blue-shifts are confined to the southern section of the LB while the
northern end of the LB develops a strong curvature with a prominent `S'-shaped hook enclosing a larger
umbral core than what was seen nearly 2.5\,hrs earlier. Although the southern section of the LB lacks the
curvature exhibited by its northern counterpart, there is a very small, although discernible, increase in 
its curvature with time. The flanks of the blue-shifted region in the southern part of the LB exhibit a
pronounced intensity  that is about 15\,\% brighter than its penumbral surroundings. A similar 
feature is seen at the northern end of the LB, where the enhanced brightness coincides with the hook-like
structure in the penumbra and lies adjacent to the blue-shifted patch. The blue-shifts reach a maximum 
value of about 850\,m\,s$^{-1}$ that occurs between 23:00--23:24 UT on 2014 March 13.

The onset of the blue-shifts is also reflected in the magnetic field wherein the field strength in the LB is
about 1500\,G and 1200\,G in the northern and southern sections of the LB, which is on average weaker than the
adjacent umbral and penumbral magnetic field by 1000\,G and 300\,G, respectively. The inclination is about 
50--60$^\circ$ in the upper and central part of the LB. In the southern section of the LB the inclination
is about 80--85$^\circ$ which is about 25$^\circ$ higher than in the neighboring penumbra. However there
are a few pixels in the southern part of the LB where the inclination is greater than 90$^\circ$.
In addition, the horizontal magnetic field is directed along the axis of the LB. Panels 32--39 
of Figure \ref{fig07r} also show that as the curvature in the northern half of the LB becomes stronger, the
values of the field strength become nearly indistinguishable from that in the adjacent umbra. The inclination
maps also show a similar characteristic where the central and southern part of the LB stands out more clearly
than its surroundings.

Rows E--G of Figure \ref{fig07r} show the response of the upper photosphere, upper chromosphere, and transition
region to the onset of the photospheric blue-shifts in the LB. Panel 11 shows a remote brightening in the 
AIA 304\,\AA~and 171\,\AA~channels close to the eastern edge of the LB. This brightening occurs at 16:49 UT,
about 13\,min after the blue-shifts appear in the photosphere. Panels 13 and 14 of Row F show the bright patch
intensifying and lying at the end of a dark filamentary structure resembling a surge. 

The AIA 1700\,\AA~images in 
Row E indicate the presence of discrete blobs of brightening on the LB (see panels 16, 26, 28) that travel  
along the LB from its southern end to the northern end. In the animation of the temporal evolution provided 
in the online material one can see that those bright grains in the 1700\,\AA~images also travel around the newly 
formed small umbral core for almost a full circle. 

The rest of the panels in Row F show that as the photospheric 
blue-shifts get stronger, the lateral span of the surges also increases 
(panel 19) extending beyond the FOV shown. The general characteristic of the surges is that they start
from the LB and extend over the eastern part of the sunspot, the filamentary ends appearing bright as also
seen in the AIA 171\,\AA~channel. The surges then fall back onto the light bridge before resuming once again.
The average lifetime of the surges based on the above characteristics is about 10--12\,min.
The animation of the AIA images shows that the surges also move bodily in a preferred direction along the 
LB. This motion starts from the southern end of the LB, coinciding with the blue-shifts, and terminates at 
the hook-like structure near the northern end of the LB, similar to the motion of the blobs seen in the 
1700\,\AA~channel. The typical length of the surges is about 5\arcsec~but they can at some instances be
as large as 18\arcsec~(panel 21). More importantly the surges do not appear elsewhere in the sunspot, 
establishing an one-to-one correspondence and causal relationship between the photospheric blue-shifts 
in the LB and the surges.

Figure \ref{fig08r} shows the LB in its decaying phase. The mid-section of the LB begins to merge eastwards
with the neighboring penumbra while the northern part of the LB starts drifting westwards. The detachment of
the northern section from the rest of the LB occurs at about 02:24 UT on 2014 March 14 (panel 59) while the
southern section too has moved closer to the penumbra. While blue-shifts are seen in the upper part of the 
LB, they are considerably weaker and the strongest blue-shifts remain concentrated at the lower end of the 
LB. However, as the LB breaks up near its central part at 02:24 UT, the blue-shifts in the southern part get
considerably weaker with time. The brightness enhancement in the continuum intensity, along the hook-like 
structure, disappears with the break-up of the LB. However the enhanced continuum intensity persists in the 
southern part of the LB lying adjacent to the blue-shifts and is observed till 06:12 UT on 2014 March 14. 
While the upper section of the LB is reduced to a chain of diffuse UDs, the southern part remains intact 
and attached to the penumbra. The last image available before the HMI data gap (panel 78) shows blue-shifts 
of about 0.17\,km\,s$^{-1}$ at the southern section of the LB. The sunspot does not split as a result
of the LB and remains as such until the early part of 2014 March 16.

The field strength maps in Row C indicate that the signature of the northern part of the LB gets weaker and
disappears at about 00:48 UT which is 1.5\,hr earlier than when the LB fragments at the center. However the
values of the field strength in the lower section of the LB, where the weak blue-shifts are seen, 
remain unchanged at 1200\,G. Similarly, the value of the inclination at that location is about 80$^\circ$.
The surges are confined to the southern section of the LB as the latter detaches from the center during its 
decay phase.

\begin{figure}[!ht]
\centerline{
\hspace{5pt}
\includegraphics[angle=0,width = \columnwidth]{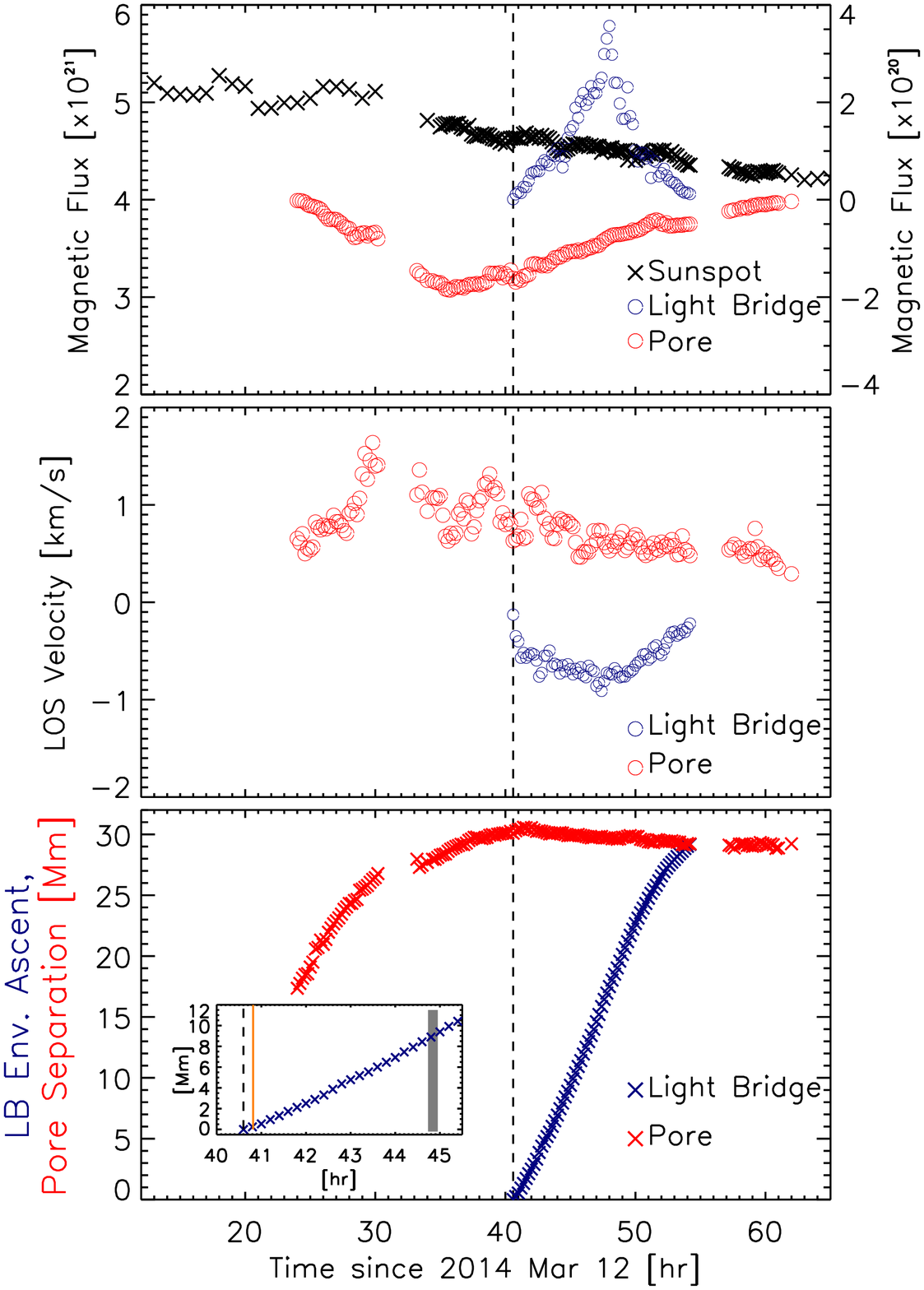}
}
\vspace{-5pt}
\caption{Top: Evolution of magnetic flux in the leading sunspot, pore, and LB. The dashed line corresponds 
to 16:36 UT on March 13 when the photospheric blue-shifts are seen in the LB. The $y$-axis on the right 
corresponds to the flux in the pore and the LB. Middle: Temporal variation of the maximum blue- and red-shifts 
in the LB and pore, respectively. Bottom: The colored $crosses$ show the height to which the envelope of the LB 
rises in the solar atmosphere (blue) and the distance of the pore separation from the sunspot center (red). The 
inset in the lower left corner shows a smaller time window of the ascent with the orange vertical line
signifying the appearance of the surges in the AIA images and the gray shaded region represent the DST observing 
window.}
\label{fig09r}
\end{figure}

The top panel of Figure \ref{fig09r} shows the evolution of magnetic flux in the leading sunspot as well 
as in the pore and the LB. The flux in the leading sunspot reduces by a factor of about 20\,\% over a period 
of 65\,hrs from a value of 5.1$\times 10^{21}$\,Mx at 13:00 UT on 2014 March 12. As seen in Section \ref{pore}, 
the flux in the pore increases rapidly over a duration of 12\,hrs from the beginning of March 13 reaching a 
maximum value of about -1.9$\times 10^{20}$\,Mx. The flux in the pore nearly remains a constant until about 
17:00 UT on March 13 after which there is a monotonic decrease in the flux. The onset of the flux reduction 
in the pore coincides with the emergence of blue-shifts in the LB which is indicated by the dashed line in 
the figure. 

The flux in the LB exhibits a sudden rise followed by a rapid decrease over a duration of 13\,hrs 
with the maximum flux of 3.6$\times 10^{20}$\,Mx at the start of March 14. The middle panel of Figure \ref{fig09r} 
shows the maximum blue- and red-shifts in the LB and pore, respectively. The blue-shifts in the LB evolve similar 
to the magnetic flux reach the peak value nearly at the same time. The red-shifts in the pore exhibit a similar 
trend with a maximum value of about 1.4\,km\,s$^{-1}$ that gradually decrease to 0.3\,km\,s$^{-1}$ towards the 
end of its lifetime. 

Using the blue-shifts in the LB, the height to which its envelope rises at a time $t_i$ was 
computed as $h(i) = h(i - 1) + (t_i - t_{i-1}) \times V_{LB}$, where $i$ and $V_{LB}$ denote the time index and 
the blue-shift in the LB, respectively. This is shown in the bottom panel of Figure \ref{fig09r} along with the 
distance of separation between the pore and the sunspot center (red crosses). The envelope reaches a height of 5\,Mm 
within 2.5\,hrs of the first appearance of the LB and a total height of 29\,Mm over its lifetime of about 13\,hrs. 
This height is also equal to the distance of separation of the pore from the sunspot center, and we find that the 
onset of the blue-shifts in the LB begin when the pore has attained this maximum separation. The inset in the 
bottom panel shows that the envelope has risen to a height of about 0.54\,Mm when the surges first appear in 
the AIA images. At the time of the DST observations, nearly 3\,hrs later, the envelope reaches a height of 
about 10\,Mm.

\begin{figure}[!h]
\centerline{
\includegraphics[angle=0,width = \columnwidth]{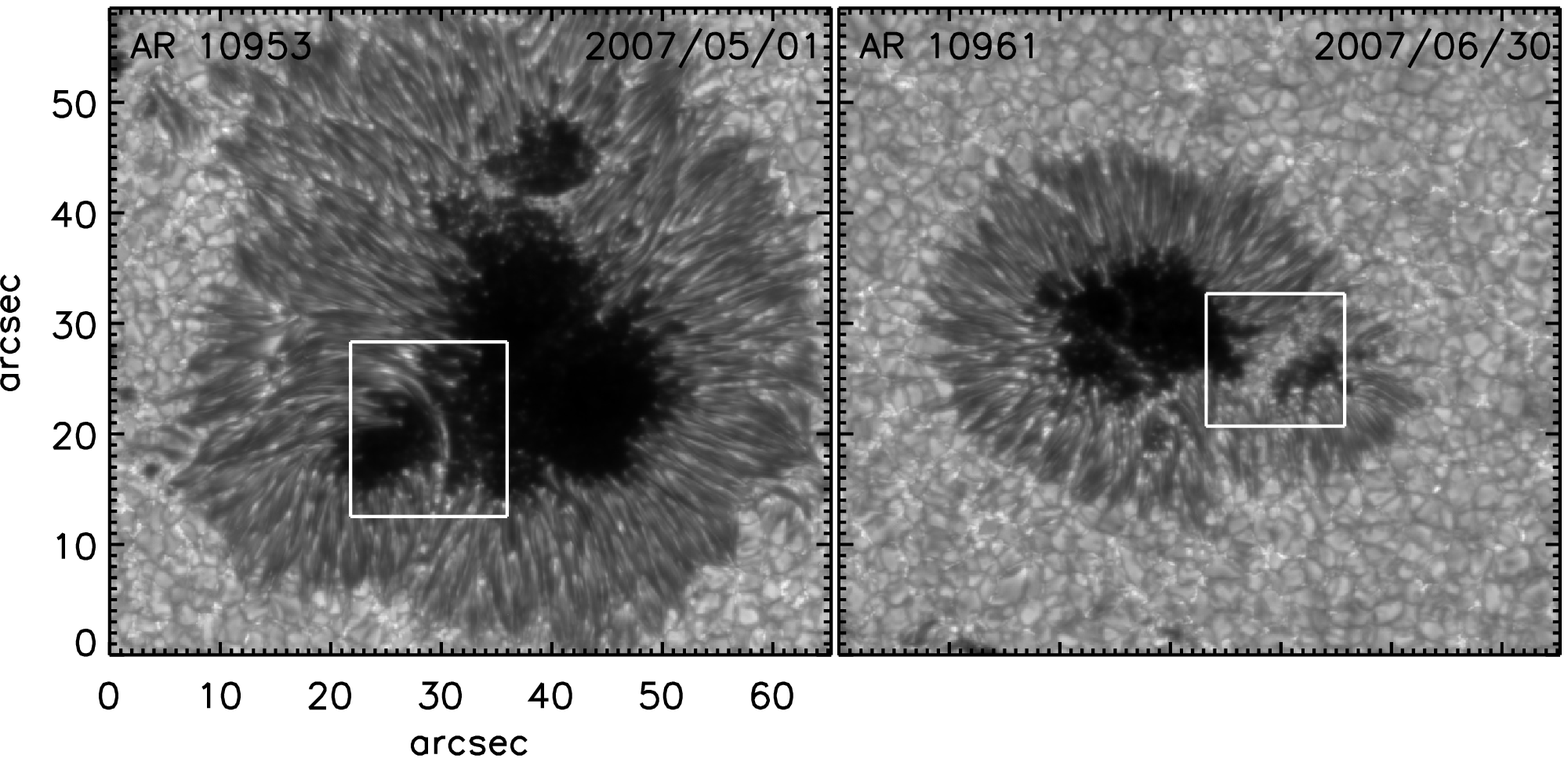}
}
\vspace{-210pt}
\centerline{
\includegraphics[angle=0,width = \columnwidth]{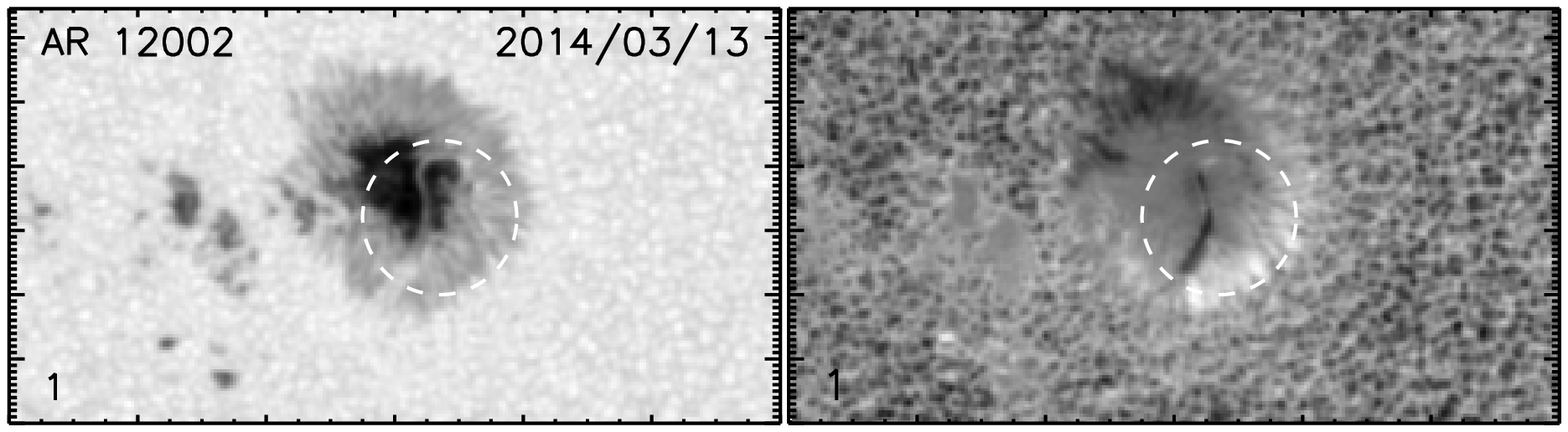}
}
\vspace{-272pt}
\centerline{
\includegraphics[angle=0,width = \columnwidth]{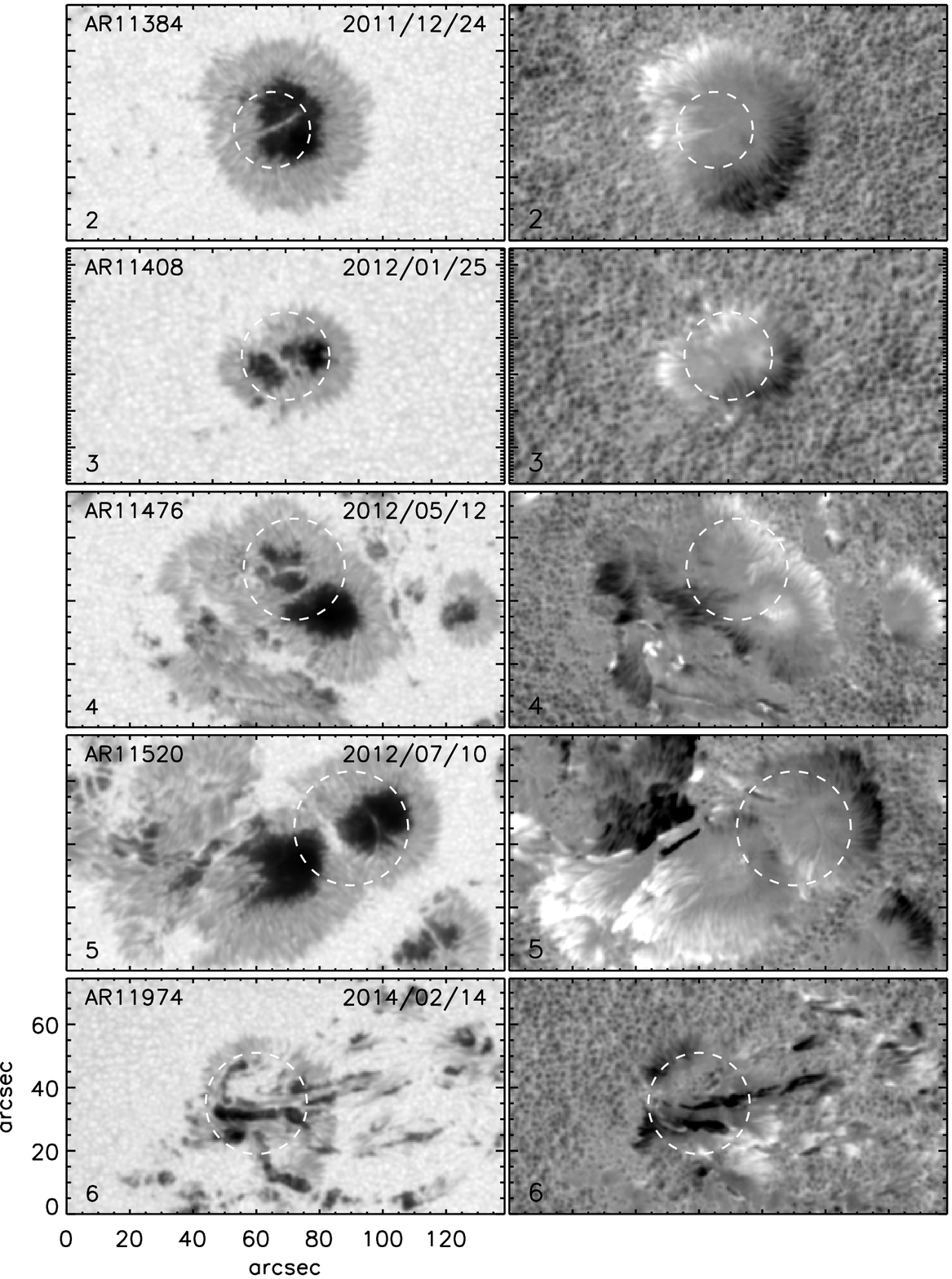}
}
\vspace{-10pt}
\caption{Examples of LBs and their associated LOS velocities. Top panel: Hinode G-band images showing 
examples of penumbral (left) and granular (right) LBs. Bottom panels: HMI Continuum intensity (left) and 
LOS velocity (right) of LBs in different ARs. The white dashed circles highlight the locations of 
the LBs. The FOV for all the bottom panels is the same, except for panels 1 and 3 
which have a FOV of 120\arcsec$\times$65\arcsec. The NOAA AR is indicated on the top left corner.}
\label{fig10r}
\end{figure}

\section{Discussion}
\label{discuss}
Sunspots LBs are regions where convective motions occur more vigorously than elsewhere 
in a sunspot. Depending on how deep the convective flows are present, the morphology of the LB can either 
be filamentary or granular as shown in the top panels of Figure \ref{fig10r} and marked with a white 
rectangle. The formation of a LB can be attributed to the 
build-up of gas pressure through convection and is often preceded by the coalescence of umbral dots, 
and intruding penumbral filaments that later develop into a larger coherent structure, resembling a LB 
\citep{2007PASJ...59S.577K}. As convection becomes more vigorous, convective cells similar to that in 
the quiet Sun develop in a granular LB, wherein upflows occur over bright cells which are flanked by 
downflows \citep{2010ApJ...718L..78R,2014A&A...568A..60L}. Filamentary LBs on the other hand either 
exhibit very weak flows or overall red-shifts (dashed circles in panels 2--5 of Figure \ref{fig10r}).
A third category of structures that extend into the umbra are referred to as umbral filaments (UFs), which 
differ from granular or filamentary LBs with respect to the morphology, temporal evolution, and magnetic
field configuration \citep{2013ApJ...770...74K,2019ApJ...880...34G}.

While the evolution of filamentary and/or granular LBs is conceived to occur within a sunspot, i.e., 
in-situ, another possible way that LBs can form in sunspots is during the emergence of an AR. 
As shown by \cite{2015ApJ...811..137T}, weakly magnetized regions in the quiet Sun can get 
squeezed by adjacent emerging magnetic flux elements as they coalesce to form sunspots. In this scenario, 
the LB can exhibit a large-scale upflow that nearly extends along its entire span (panel 6 of Figure \ref{fig10r}). 
These observations of \cite{2015ApJ...811..137T} are consistent with results obtained from radiative MHD 
simulations of flux emergence as described in \cite{2008ApJ...687.1373C,2010ApJ...720..233C}. 

\begin{figure*}
\begin{minipage}{8cm}
\resizebox{8cm}{!}{\includegraphics{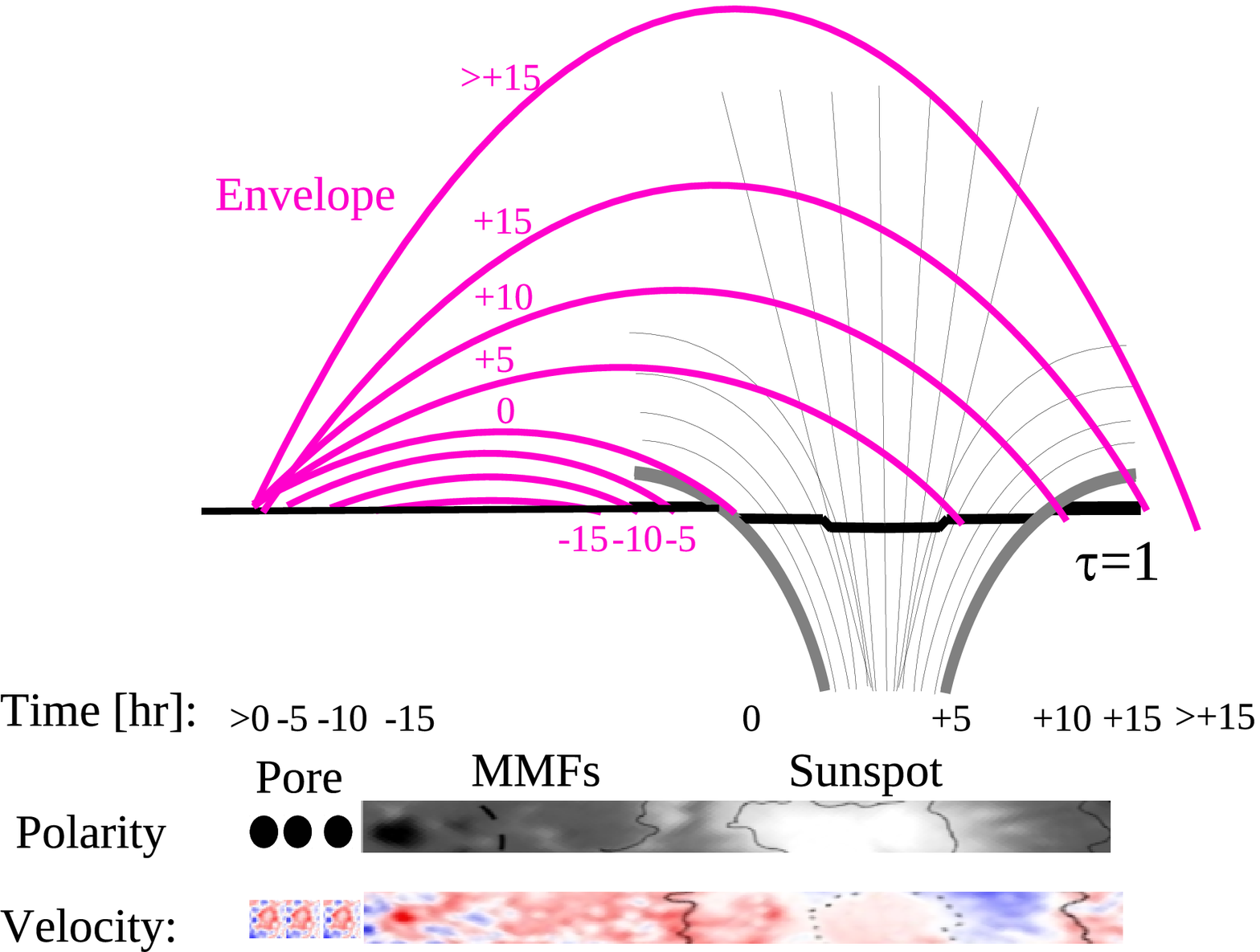}}
\end{minipage}\hspace*{1cm}
\begin{minipage}{8cm}
\vspace*{.0cm}
\resizebox{8cm}{!}{\includegraphics{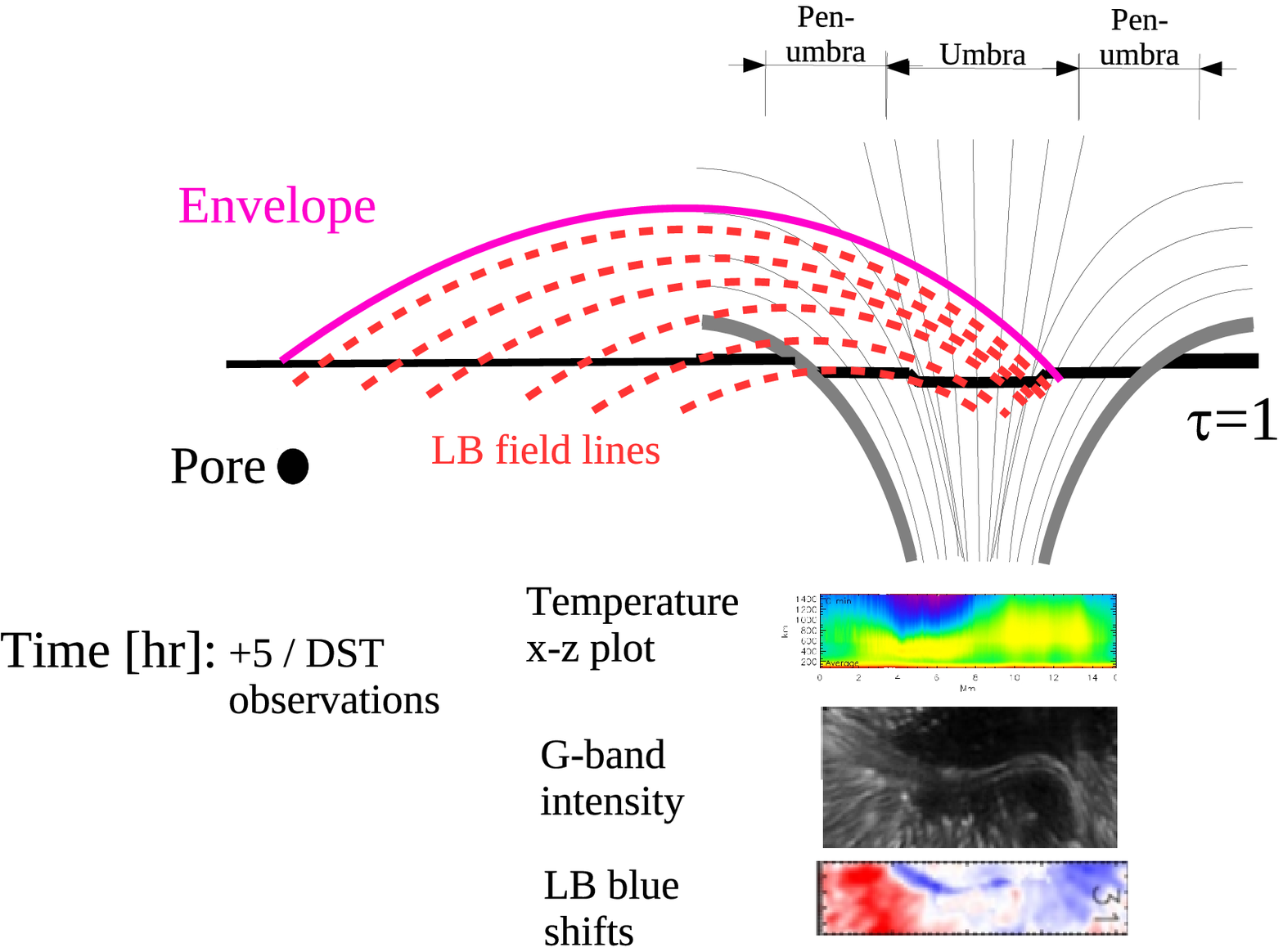}}
\end{minipage}
\caption{Sketch of the temporal evolution and the LB structure at the time of the DST observations on March 13, 
UT 21:00. South is at the left. A time of $t=0$\,hr marks the first appearance of the blue shifts at the outer 
penumbral boundary of the sunspot, whereas the DST observations were at about $t=+5$\,hr. Left panel: possible 
magnetic topology during the magnetic flux emergence. Thick pink lines indicate the envelope of the emerging 
flux, i.e., the highest field line. The thick grey line marks the outer end of the sunspot fields. Thin black 
lines indicate individual field lines in the penumbra and umbra. The thick black line marks the height of the 
$\tau=1$ layer. The location of the pore is indicated for $t=-15$ to $>0$\,hrs. Right panel: possible magnetic 
field topology during the DST observations at $t=+5$\,hrs (red dashed lines). At the southern end of the LB 
the field lines have a higher inclination than the surrounding penumbra.}
\label{figsketch}
\end{figure*}

In this article, the LB forms in a regular, well-developed sunspot with a simple magnetic configuration. 
The LB is a transient structure that survives for about 13\,hrs. The most outstanding feature of 
our observation is that the LB exhibits blue-shifts nearly along its entire length with values of up to 
0.85\,km\,s$^{-1}$ (panel 1 of Figure \ref{fig10r}). The blue-shifted structure extends across one part 
of the umbral core and into the limb-side penumbra which is dominated by red-shifts associated with the normal 
Evershed flow. 
We trace the formation of the LB to an emergence event nearly 
17\,hrs earlier wherein one leg of the emergent structure is located near the outer penumbra of the sunspot 
and the other leg coincides with a pore in the quiet Sun just outside the visible sunspot boundary. 
As the pore recedes from the sunspot, a filamentary channel develops in the penumbra extending from the 
penumbra-quiet Sun boundary to the umbra-penumbra boundary where the LB consequently forms.

The highly inclined LB with a relatively weaker field strength, compared to the umbra, renders it brighter 
than the rest of the sunspot as well as the quiet Sun from the upper photosphere to the lower chromosphere. 
The high-resolution chromospheric observations from the DST reveal that the thermal distribution across 
the LB is quite complex, with the boundary enclosing the hot structure changing dramatically along its length, 
across its width, as well as in time, over a duration as short as 11\,min. 
We also find evidence from the thermal stratification that certain sections of the LB are 
disconnected from the photosphere appearing nearly isolated at mid chromospheric heights. These support the 
idea of a rising, hot structure wherein different sections are elevated at different heights.
The observed inhomogeneous thermal structure of the LB could be attributed to the following.
At the photosphere the LB comprises weak, and highly inclined magnetic fields. However, the vertical extent of the 
magnetic field as well as that within the LB could be such that there may arise small-scale electric currents 
that produce local enhancements in temperature. Additionally, the LB appears bright in the photospheric 
continuum which would imply that the structure is hot, resulting from the emergence of sub-photospheric 
flux. There could be an additional pressure gradient along the length of the LB if one traces the blue-shifted 
patches or blobs that move from the southern to the northern end of the LB. A flow along such a structure 
elevated to different heights would disperse the thermal flux non-uniformly that would render a complex 3D 
temperature distribution in the LB.

Figure \ref{figsketch} shows a sketch of the possible magnetic topology during the magnetic flux emergence 
and the LB existence. The distance of the pore from the sunspot increases from 17\,Mm to 30\,Mm from $t=-15$ 
to $0$\,hrs and remains at that value afterwards. The integrated height corresponding to the maximal blue 
shift velocity in the LB increases from 0\,Mm at $t=0$\,hrs to 30\,Mms at $t=+15$\,hrs. The MMFs between 
the pore and the sunspot partly have the same polarity as the sunspot. The pore shows persistent red shifts. 
The LB disappears about 13 hrs after its emergence. 

At the time of the DST observations, the envelope has 
risen to about 10\,Mm. The thermal inversion traces increased temperatures at about 0.5\,Mm at the southern 
end of the LB and at 0.8\,Mm at the northern end. The magnetic field inclination at the southern end, where 
the G-band image shows penumbral grains at the outer penumbral boundary, could be inclined towards the umbra 
opposite to the surrounding magnetic field lines. Those field lines might be able to sustain an Evershed flow 
in the opposite direction towards the umbra. The blue shifts persist along the length of the LB up to the point 
where it turns by 90$^\circ$ out of the plane of the figure in the `S'-shaped hook.
 
We now address the interaction of the emergent structure, corresponding to the LB, with the overlying umbral magnetic 
field. The appearance of the photospheric blue-shifts in the LB is accompanied by recurrent, and highly dynamic, 
surges in the chromosphere and transition region that start about 13\,min later. Assuming an ascent speed of 
1.0\,km\,s$^{-1}$ (the maximum blue-shift is about 0.85\,km\,s$^{-1}$), the loop would reach lower chromospheric 
heights of 0.78\,Mm in the above time span. 
Furthermore, the surges persist till the lifetime of the LB, and cease completely when the LB dissolves. 
Thus our observations establish the causal nature of the surges with the photospheric blue-shifts in the LB 
and explain the recurrent behavior of the surges as a direct consequence of a large-scale, nearly horizontal, emergent 
structure within a sunspot. The physical attributes of the surges are in good agreement with \cite{2016A&A...590A..57R}
and the bodily motion can be due to the emergence of the LB that progresses from the south to the north. 
We find that the LB as a whole is hotter than the umbra along all of its length, while the surges are much more intermittent 
in both space and time. The increased chromospheric temperature of the LB thus is more likely to be due to ohmic dissipation 
of electric currents \citep[e.g.,][]{2008ApJ...686L..45T} caused by the large magnetic field gradient to its surroundings rather 
than the more transient surges.
While earlier studies suggest magnetic reconnection as a mechanism to drive the surges 
\citep{2016A&A...590A..57R,2018ApJ...854...92T},
determining the exact mechanism for the surges is beyond the scope of this article. 

Apart from the surges, we do not
observe any flaring activity associated after the onset of the blue-shifts in the LB during the emergence process, 
possibly due to the low flux content in the structure accounting for just 4\,\% of the sunspot. Transient jets, are 
however seen during the early stages of the pore's emergence as a result of small-scale magnetic reconnection with 
opposite polarity patches being driven by the moat flow \citep{2015A&A...583A.110M}. 
The dissolution of the LB is possibly due to counter-clockwise rotation of the sunspot, wherein the northern and 
southern ends of the LB move in opposite directions and cause the LB to eventually break up near its mid-section.

The physical characteristics of the LB described in this article suggest that it is closer to an UF rather than a 
filamentary or granular LB.  There are however, distinct differences with the UFs reported by 
\citet{2013ApJ...770...74K} and \citet{2019ApJ...880...34G}. According to \citet{2013ApJ...770...74K}, UFs are separate 
flux bundles that possibly connect the umbra to a network region outside the sunspot, wherein a siphon flow develops 
driving a counter-Evershed flow from outside the spot into the umbra. However, in their observations, the LB exhibits 
red-shifts with blue-shifts confined exclusively to the penumbral section of the sunspot similar to the results of 
\citet{2019ApJ...880...34G}. 
In contrast, the LB described here has a persistent blue-shift all along its length which is compatible with an emergent 
structure, not a siphon flow. The outer footpoint of our emergent structure coincides with a pore that is red-shifted, 
whereas a siphon flow ought to produce blue-shifts at that location. Furthermore, the UF described by 
\citet{2019ApJ...880...34G} shows that the UF as well as the blue-shifted penumbral region have an opposite polarity 
as the sunspot, whereas the magnetic field in the LB described here has the same polarity as the sunspot, albeit highly 
inclined. The southern section of the LB near the sunspot-Quiet Sun boundary coincides with bright penumbral grains 
which is not seen in \cite{2013ApJ...770...74K}. On the other hand, while this feature is present in 
\citet{2019ApJ...880...34G}, it is located further inwards in the penumbra. 

\citet{2013ApJ...770...74K} also suggest that the brightness of UFs in the chromosphere is related to dissipation 
that occurs in the upper photospheric or lower chromospheric heights which is compatible with the temperature 
enhancements we detect between 400--800\,km. However, these enhancements do not originate from flares or other 
energetic phenomena as seen by \citet{2013ApJ...770...74K}.

Based on the above, one can distinguish LBs from UFs using the following criteria. UFs lack multiple convective cells 
and/or dark lanes or striations. In comparison to LBs, UFs are highly elongated, with a small radius of curvature and 
are short-lived. In addition, if the UFs exhibit blue-shifts all along their length, it is must signify emergence.
While our observations indicate that the structure in the sunspot umbra resembles a generic LB, it is however, part 
of a large-scale emergence event initiated just outside the sunspot. The magnetic association of a LB to an external 
structure is also observed in AR filaments that terminate close to the outer ends of sunspot LBs 
\citep{2010ApJ...711.1057G,2011ApJ...738...83S,2015A&A...583A.110M}.

Our results suggest that the sunspot magnetic field is more `gappy' at sub-photospheric layers to facilitate 
flux emergence as shown by \cite{2014A&A...562A.110L} where the rear half of a leading sunspot was connected 
to the following polarity of the AR while the front half was rooted in the preceding active region. As a result 
of this unique magnetic connectivity, the leading sunspot eventually split into two, nearly equal halves, with 
the front half separating and moving rapidly away from its twin. The delay in the formation of the LB nearly 
17\,hr after pore's emergence shows that magnetic pressure is quite strong to inhibit the rise, initially, 
but also fragmented at the sub-surface to allow the structure to eventually rise. The presence of a penumbral 
intrusion prior to the LB formation possibly indicate a weakening of the umbral magnetic field which is similar 
to that shown by \cite{2007PASJ...59S.577K}. During its lifetime, we find the LB does not develop sufficiently 
to exhibit a granular morphology. Its physical properties and the large-scale external association to the pore 
render it atypical to what is generically referred to as LBs.

Our observation of an elongated LB in the umbra suggests that a sunspot should harbor sub-photospheric 
structures which can emerge under suitable conditions on various spatial scales, as in our case, comparable to the 
size of the sunspot. Such emerging magnetic flux can interact with the pre-existing structure leading to surges and 
possibly flares, presumably through magnetic reconnection. The flux emergence did not, however, destabilize the 
sunspot nor disrupt its internal structure enough to trigger its decay. An extended study on additional cases of 
umbral flux emergence on different spatial scales might reveal any eventual impact on the lifetime of sunspots 
in contrast to sunspot decay due to the erosion at the outer penumbral boundary \citep{1964ApJ...140.1120S,2015ApJ...800..130L}.

\section{Conclusions}
\label{conclu}
The LB under investigation formed as a result of large-scale flux emergence wherein one leg of the emergent structure 
is rooted in a pore located in the quiet Sun, while the LB forms the flatter segment of the loop. The emergence is 
characterized by conspicuous blue-shifts in the LB whose southern end runs well into the limb-side penumbra that is 
red-shifted. The receding motion of the pore away from the sunspot is accompanied by a rise of the emergent LB in 
the solar atmosphere, which produces highly dynamic surges in the chromosphere and transition region. While the LB is hotter
than the adjacent umbra by about 600--800\,K, various segments of the LB are elevated at heights that vary between 400 to 700\,km. 
The LB only survives for about 13\,hrs, while the highest field lines related to the flux emergence could reach a height
of about 29\,Mm in the solar atmosphere. Although the LB has a morphology similar to a penumbral 
intrusion in the umbra, its physical properties and its association to the pore make it an atypical LB. We suggest that the 
presence of large-scale persistent blue-shifts in LBs can be used to distinguish structures that are driven by in-situ 
magneto-convection from those caused by large-scale flux emergence in a sunspot.

\acknowledgments
The Dunn Solar Telescope at Sacramento Peak/NM was 
operated by the National Solar Observatory (NSO). NSO is operated by the Association of Universities for Research 
in Astronomy (AURA), Inc. under cooperative agreement with the National Science Foundation (NSF). HMI data are 
courtesy of NASA/SDO and the HMI science team.  They are provided by the Joint Science Operations Center -- Science
Data Processing at Stanford University. IBIS has been designed and constructed by the INAF/Osservatorio Astrofisico 
di Arcetri with contributions from the Universit{\`a} di Firenze, the Universit{\`a}di Roma Tor Vergata, and upgraded
with further contributions from NSO and Queens University Belfast. This work was supported through NSF grant
AGS-1413686. We thank the anonymous referee for evaluating our manuscript and providing useful comments.


\end{document}